\documentclass[twocolumn,journal]{IEEEtran}
\makeatletter
\long\def\@makecaption#1#2{\ifx\@captype\@IEEEtablestring%
\footnotesize\begin{center}{\normalfont\footnotesize #1}\\
{\normalfont\footnotesize\scshape #2}\end{center}%
\@IEEEtablecaptionsepspace
\else
\@IEEEfigurecaptionsepspace
\setbox\@tempboxa\hbox{\normalfont\footnotesize {#1.}~~ #2}%
\ifdim \wd\@tempboxa >\hsize%
\setbox\@tempboxa\hbox{\normalfont\footnotesize {#1.}~~ }%
\parbox[t]{\hsize}{\normalfont\footnotesize \noindent\unhbox\@tempboxa#2}%
\else
\hbox to\hsize{\normalfont\footnotesize\hfil\box\@tempboxa\hfil}\fi\fi}
\makeatother

\usepackage{multirow}
\usepackage[dvips]{graphicx}
\usepackage{bbm}

\usepackage{amsmath}
\usepackage{amsxtra}
\usepackage{bbold}
\usepackage{algorithm}
\usepackage[noend]{algpseudocode}
\usepackage{threeparttable}
\usepackage{cite}
\usepackage{bm}
\usepackage{bbding}
\usepackage{pifont}
\usepackage{wasysym}
\usepackage{amssymb}

\usepackage{epstopdf}
\usepackage{eso-pic}
\usepackage{adjustbox}
\usepackage{amsmath}
\DeclareMathOperator*{\argmax}{arg\,max}
\DeclareMathOperator*{\argmin}{arg\,min}
\newcommand{\quotes}[1]{``#1''}
\usepackage{booktabs}
\usepackage{flushend}
\usepackage{soul}
\usepackage[T1]{fontenc}
\usepackage[utf8]{inputenc}
\usepackage{pict2e,picture}
\newsavebox\CBox
\newlength\CLength
\def\Circled#1{\sbox\CBox{#1}%
  \ifdim\wd\CBox>\ht\CBox \CLength=\wd\CBox\else\CLength=\ht\CBox\fi
    \makebox[1.2\CLength]{\makebox(0,1.2\CLength){\put(0,0){\circle{1.2\CLength}}}%
    \makebox(0,1.2\CLength){\put(-.5\wd\CBox,0){#1}}}}

\usepackage{tikz}
\newcommand*\circled[1]{\tikz[baseline=(char.base)]{
            \node[shape=circle,draw,inner sep=1.25pt] (char) {#1};}}
\ifCLASSINFOpdf
\else
\fi

\begin{document}
%\title{Exploiting Mismatched Language Resources  for Unsupervised Subword Modeling}
\title{Exploiting Cross-Lingual Speaker and Phonetic Diversity for Unsupervised Subword Modeling}
% \title{Unsupervised Subword Modeling of Zero-Resource Languages with Cross-Lingual Knowledge Transfer}

\author{Siyuan~Feng~
        and~Tan~Lee% <-this % stops a space
\thanks{S. Feng and T. Lee are with the Department
of Electronic Engineering, The Chinese University of Hong Kong, Hong Kong SAR, China (e-mail: siyuanfeng@link.cuhk.edu.hk; tanlee@ee.cuhk.edu.hk).}% <-this % stops a space

\thanks{This research is partially supported by a GRF project grant (Ref: CUHK 14227216) from Hong Kong Research Grants Council.}}


\markboth{}%
{Shell \MakeLowercase{\textit{et al.}}: Bare Demo of IEEEtran.cls for IEEE Journals}



\maketitle

\begin{abstract}
%NOT YET MODIFIED.
% Unsupervised acoustic modeling aims at modeling speech at subword or word level, assuming only untranscribed speech data are available. 
% This is often referred to as the zero-resource problem. 
This research addresses the problem of acoustic modeling of low-resource languages for which transcribed training data is absent. The goal is to learn robust frame-level feature representations that can be used to identify and distinguish subword-level speech units. The proposed feature representations comprise various types of multilingual bottleneck features (BNFs) that are obtained via multi-task learning of deep neural networks (MTL-DNN). One of the key problems is how to acquire high-quality frame labels for untranscribed training data to facilitate supervised DNN training. It is shown that learning of robust BNF representations can be achieved by effectively leveraging transcribed speech data and well-trained automatic speech recognition (ASR) systems from one or more out-of-domain (resource-rich) languages. Out-of-domain ASR systems can be applied to perform speaker adaptation with untranscribed training data of the target language, and to decode the training speech into frame-level labels for DNN training. It is also found that better frame labels can be generated by considering temporal dependency in speech when performing frame clustering. The proposed methods of feature learning are evaluated on the standard task of unsupervised subword modeling in Track 1 of the ZeroSpeech 2017 Challenge. The best performance achieved by our system is $9.7\%$ in terms of across-speaker triphone minimal-pair ABX error rate, which is comparable to the best systems reported recently. Lastly, our investigation reveals that the closeness between target languages and out-of-domain languages and the amount of available training data for individual target languages could have significant impact on the goodness of learned features.

\end{abstract}


% \fi
\IEEEpeerreviewmaketitle

\section{Introduction}
\IEEEPARstart{S}{tate-of-the-art} automatic speech recognition (ASR) systems have demonstrated fairly impressive performance in terms of word accuracy \cite{Saon2017,Hori2017}. This is mainly attributed to the advances of deep neural network (DNN) based acoustic models (AMs) and language models (LMs) \cite{hinton2012deep,ragni2016multi}. Typically a well-trained DNN-based AM requires hundreds to thousands of hours of transcribed speech. As a matter of fact, high-performance ASR systems are available only for major languages \cite{shibata2017composite}. Even for resource-rich languages, preparing transcriptions for available training data is a time-consuming task that involves considerable human effort. For many languages in the world, very little or no transcribed speech is available \cite{dunbar2017zero}, and conventional acoustic modeling techniques are simply not applicable.

Unsupervised speech modeling is the task of building sub-word or word-level AMs, when only untranscribed speech are available for training \cite{glass2012towards,kamper2015fully,versteegh2015zero}. This is also known as the \textit{zero-resource} problem, which has attracted increasing research interest in recent years.
% Considering it as a very difficult problem, there have been increasing  research interests in zero-resource speech modeling in recent years.
%Zero-resource speech modeling has been gaining much attention, especially  in the recent years. 
The Zero Resource Speech Challenge 2015 (ZeroSpeech 2015) \cite{versteegh2015zero} and 2017 (ZeroSpeech 2017) \cite{dunbar2017zero} precisely focused on unsupervised speech modeling. ZeroSpeech 2017 was organized to tackle two sub-problems, namely \textit{unsupervised subword modeling} (Track 1) and \textit{spoken term discovery (STD)} (Track 2). The present study addresses the Track 1 problem and aims to learn frame-level feature representation that is effective in identifying and discriminating subword-level units and robust to irrelevant factors, e.g., speaker and/or channel variation, emotion, etc. Robust feature representations obtained by learning from data have been found to be preferable to conventional spectral features like Mel-frequency cepstral coefficients (MFCCs) for downstream applications  \cite{Chen+2016,yuan2017pairwise}.

%  incorporating the idea of adversarial learning \cite{ganin2015unsupervised}, which was firstly applied to noise robust ASR \cite{shinohara2016adversarial}, in order to extract features for zero-resource languages that are invariant to speakers while discriminative to phonemes.

% There are two representative frameworks of frame-level feature representation learning in the unsupervised scenario, namely, deep learning (DL)-based and non DL-based. 

% {\color{blue}[There is no separate Literature review section.]} 
% In previous studies on unsupervised subword modeling, d
DNN models are commonly adopted in frame-level feature learning for unsupervised subword modeling. A DNN model is typically trained using available speech data. The learned features are obtained either from a designated low-dimension hidden layer of the DNN, known as the bottleneck features (BNFs) \cite{chen2017multilingual}, or from the softmax output layer, known as the posterior features or posteriorgram \cite{ansari2017deep}. To facilitate supervised training of the DNN, target labels of training speech are needed.
In zero-resource scenarios, the key problem is how to generate informative frame-level labels in the absence of speech transcription. One of the possible approaches is based on unsupervised clustering of training speech. The frame-level cluster indices can be used as target labels for DNN training \cite{chen2017multilingual,yuan2017pairwise,ansari2017deep}. 
Another approach seeks to use pre-trained out-of-domain ASR systems to tokenize untranscribed in-domain speech and hence each frame is assigned with an ASR senone label \cite{shibata2017composite,feng2018exploiting}.
% {\color{cyan}, or to apply transfer learning techniques to generate target labels[*i suggest to avoid mentioning transfer learning here, and move to section II*]}.
Fully unsupervised \cite{ansari2017deep} or weakly supervised \cite{synnaeve2014weakly,kamper2015unsupervised,hermann2018multilingual_journal} methods for DNN training were also reported in the research on acoustic modeling for low-resource languages.

% For unsupervised models,  target labels are not required during model training. 
% % AEs generate non-linear embeddings of input speech while also learning a reconstruction from the embeddings \cite{bengio2009learning}. 
% % ``Weakly supervised'' or ``pair-wise supervised'' networks were also studied in the concered task \cite{synnaeve2014weakly,kamper2015unsupervised,hermann2018multilingual_journal}. 
% For weakly supervised (also known as pair-wise supervised) models, while transcriptions for target speech are unavailable, same-different pairing of speech segments is assumed known or at least readily available, e.g. from unsupervised term discovery (UTD) results. Pair-wise information  serves as weak supervision, which has been proved to improve subword discriminability compared with unsupervised training paradigms \cite{renshaw2015comparison}. 
 
The present study adopts the general framework of supervised DNN training for the purpose of extracting BNF as the learned feature representation.
%Each frame is represented by an HMM state posterior vector.
%The inputs to the DNN are spectral features and targets for DNN training could be
We attempt to improve the efficacy and performance of learned features along two directions. First, advanced unsupervised acoustic modeling techniques are explored to generate initial frame-level labels for supervised DNN training. Second, speaker adaptation techniques are applied to make input speech features more robust to speaker variation.

% The idea of frame-level label acquisition at the very beginning of the modeling procedure was shared across different studies
%  \cite{heck2017feature,chen2017multilingual,ansari2017deep}. The intuition behind is that, to apply conventional well-established acoustic modeling techniques to the zero-resource scenario, a natural way is to find some kinds of frame labels as initial supervision.

 % {\color{cyan}To this end, some researchers exploit   phone recognizers for resource-rich languages to decode and generate frame-level labels for unsupervised speech of target zero-resource languages \cite{shibata2017composite,feng2018exploiting}. One shortcoming of this approach is that, the mismatch between source and target domain, e.g. language, speaker and channel, may probably lead to inaccurate frame labels.}
Dirichlet process Gaussian mixture model (DPGMM) is commonly used for clustering of unlabelled speech frames \cite{chang2013parallel}. It demonstrated superior performance on the tasks in ZeroSpeech Challenges \cite{chen2015parallel,heck2017feature}. However, DPGMM clustering, as well as other conventional clustering algorithms like $k$-means \cite{manenti2017unsupervised} and GMM \cite{ansari2017deep}, assumes that neighboring speech frames are independent of each other. This is obviously not in accordance with the nature of speech. 
% {\color{blue} 
To address this limitation, a full-fledged Gaussian mixture model-hidden Markov model (GMM-HMM) AM is trained to capture contextual information in speech. The transcriptions required for GMM-HMM training are initialized via DPGMM clustering. Following the terminology in \cite{Heck2016iterative}, this model is referred to as \emph{DPGMM-HMM}.
We use the DPGMM-HMM AM to generate frame-level labels to support BNF representation learning. In \cite{Heck2016iterative}, a similar approach was adopted for learning feature-space maximum likelihood linear regression (fMLLR) and posteriorgram features.
% }

In unsupervised subword modeling, the outcome of frame clustering ideally comprises a set of clusters that correspond to phoneme-related speech units. The underlying assumption is that speech frames identified as the same phoneme should have homogeneous acoustic properties.
In practice, speaker and environment variations would inevitably impact the reliability of frame clustering results.
Our preliminary experiments showed that applying DPGMM typically results in an excessive number of fine-grained clusters. Similar observations were reported in \cite{heck2018dirichlet,wu2018optimizing}. These over-fragmented clusters may adversely affect the effectiveness of unsupervised speech modeling. In this work we develop and apply a new algorithm to filter out infrequent labels in DPGMM clustering results, and experimentally validate its effectiveness.

% {\color{blue}
In addition to the DPGMM-HMM labels, a different type of frame labels can be obtained using one or more out-of-domain ASR systems \cite{shibata2017composite,feng2018exploiting}. While the DPGMM-HMM frame labels incorporate statistical information of the acoustic properties of target speech, the ASR senone labels leverage the phonetic information acquired from out-of-domain languages. We propose to exploit their complementarity in DNN based feature learning by applying the multi-task learning strategy \cite{caruana1998multitask}.
% }

Numerous studies have demonstrated the benefit of applying speaker adaptation on input features for unsupervised subword modeling \cite{chen2017multilingual,Heck+2016}. In the present study, we propose to exploit cross-lingual speech data in fMLLR-based speaker adaptation. Specifically, transcribed speech from a resource-rich language is used to train an out-of-domain ASR system. This system is then applied to the zero-resource target languages for estimating linear discriminant analysis (LDA), maximum likelihood linear transform (MLLT) and fMLLR transforms on conventional spectral features. We advocate that this approach is effective and practically desirable as transcribed speech data of resource-rich languages are relatively easy to access.

The remainder of this paper is organized as follows.  Section \ref{sec:related_works} provides a review of related works on unsupervised subword modeling with untranscribed speech.
%In Section \ref{sec_related_works}, we provide a brief overview on the previous works related to unsupervised feature representation learning for zero-resource languages. 
In Section \ref{sec:framework}, we provide detailed description on the proposed approaches to feature learning.  Section \ref{sec:exp} introduces experimental design on ZeroSpeech 2017 development data. Section \ref{sec:results_and_analyses} discusses and analyzes experimental results. Section \ref{sec:conclusion} gives the conclusions.
\label{sec:intro}

\section{Related works}
\label{sec:related_works}
% List related works with some ordered categories here.
\subsection{Deep learning approaches to unsupervised subword modeling}

A variety of DNN models have been investigated towards unsupervised subword modeling. They include multi-layer perceptron (MLP) \cite{chen2017multilingual}, auto-encoder (AE) \cite{ansari2017deep}, correspondence AE (cAE), denoising AE (dAE) \cite{renshaw2015comparison}, variational AE (VAE) \cite{chorowski2019unsupervised} and siamese network \cite{yuan2017extracting}. In terms of training strategies, these models can be classified into three categories, namely, supervised (MLP), unsupervised (AE, VAE, dAE) and weakly/pair-wise supervised (cAE, siamese network). Supervised DNN training requires frame-level labels for all training speech, which could be obtained either via a clustering process or exploiting out-of-domain resources.
% as mentioned in Section \ref{sec:intro}. 
In \cite{chen2017multilingual,yuan2017pairwise}, DPGMM clustering was performed on conventional short-time spectral features of target speech, followed by multilingual DNN training to obtain the BNF representation. In \cite{ansari2017deep}, GMM-universal background model (GMM-UBM) was used to generate frame labels. A DNN was trained using these labels to generate BNF or posteriorgram representation. In \cite{shibata2017composite,feng2018exploiting}, language-mismatched ASR systems were utilized to decode the target speech, and frame labels were generated from the ASR decoding lattices.
% {\color{blue} 
In \cite{chen2017multitask_jstsp}, BNF representation was generated by applying multi-task learning with both in-domain and out-of-domain data \cite{caruana1998multitask}. The frame labels for out-of-domain data were obtained by HMM forced alignment, while the labels for in-domain data were from DPGMM clustering \cite{chen2017multilingual}.
% }
% {\color{cyan}
% Out-of-domain resources could also be exploited in the \emph{transfer learning} manner \cite{shibata2017composite,tsuchiya2018speaker,feng2018exploiting}, i.e.,
In \cite{shibata2017composite,tsuchiya2018speaker,feng2018exploiting}, a DNN AM 
% for an out-of-domain language 
was trained with transcribed data of an out-of-domain language, and used to 
extract BNFs or posteriorgrams from target speech.
% }
% In this approach, supervised DNN training is carried out with out-of-domain transcribed data.}

Unsupervised DNN training does not require any kind of target labels. For example, an AE model generates non-linear embeddings of input speech and meanwhile learn to reconstruct the same speech from the embeddings.
% The objective function of an AE is  mean square error of input features and their corresponding AE output features. 
Recently, weakly-supervised model training is studied extensively \cite{synnaeve2014weakly,kamper2015unsupervised,hermann2018multilingual_journal}. 
In the cAE model \cite{renshaw2015comparison}, a pair of speech segments that contain the same linguistic unit (word or subword) are used as the input and output for training, with the objective of minimizing the reconstruction error. In a siamese network, the input comprises two speech segments. The network is trained to determine whether the segments are from the same linguistic unit or not. These models were shown to achieve better performance than unsupervised models \cite{renshaw2015comparison}. However, for zero-resource languages, such pair-wise knowledge may not be directly available.
% Applying weak supervision has been proved able to improve subword discriminability compared with unsupervised network training frameworks \cite{renshaw2015comparison}.
% To train the DNN for unsupervised subword modeling, 

% \subsection{Input features for unsupervised subword modeling}

% } 

\subsection{Unsupervised subword modeling without using DNN}

There were numerous studies on unsupervised subword modeling without involving deep learning models. In these studies, clustering of short-time frame features is an important first step.
% researchers usually perform clustering towards speech data to obtain a certain number of clusters. 
After frame clustering, each cluster is represented by a learned probability distribution, and the cluster posteriorgram can be regarded as the learned representation for subword modeling.
% for subword discriminability evaluation.
% at the first stage, either at frame-level or segment-level. After clustering, target zero-resource speech is represented by a cluster posteriorgram.
Frame clustering could be done straightforwardly by applying
 $k$-means \cite{manenti2017unsupervised},  GMM \cite{ansari2017unsupervised} and DPGMM \cite{chen2015parallel} algorithms. 
%  {\color{cyan}[may we not start a new paragraph below?]}
In \cite{chen2015parallel}, DPGMM clustering was applied to a zero-resource target language. An extension of this approach was reported in \cite{heck2017feature}, where clustering was performed with fMLLR-based speaker-adapted features. 
% The work in \cite{heck2017feature} extended \cite{chen2015parallel} by clustering fMLLR features of target speech frames, and achieved improved performance. 
% The fMLLRs are estimated by performing a first-pass frame clustering to obtain initial labels as pseudo transcription, followed by conventional GMM-HMM acoustic modeling with speaker adaptive training. 
In \cite{ansari2017unsupervised}, GMM posteriorgram and HMM posteriorgram were compared, where the HMM was trained based on GMM-UBM clustering results. 
% In \cite{manenti2017unsupervised}, $k$-means was applied 
% Results in \cite{ansari2017unsupervised} show that HMM posteriorgram outperforms GMM posteriorgram in unsuperivsed subword modeling task.

% The work in \cite{ansari2017unsupervised} further extended GMM posteriorgram to hidden Markov model (HMM) posteriorgram by performing supervised GMM-HMM acoustic modeling, with transcription obtained by GMM initialization.

To better retain temporal dependency in speech, frame clustering can be embodied in segment level. Initial segmentation of speech utterances could be obtained by hierarchical agglomerative clustering \cite{QiaoShimomuraMinematsu2008}, or using language-mismatched phone recognizers \cite{feng2016exploit,sung2018unsuperivsed}. Subsequently a fixed-length feature vector is derived to represent each speech segment. Clustering of segment-level feature vectors was tackled using a range of algorithms, including vector quantization (VQ) \cite{LeeSoongJuang}, segmental GMM (SGMM) \cite{GishNg1993}, spectral clustering \cite{I3EWang} and graph clustering \cite{Bhati2017}. In \cite{kamper2016unsupervised}, segmentation and clustering were integrated as a jointly optimized process.

The present study is on one hand largely based on DNN modeling of speech, and on the other hand incorporates the ideas of frame clustering (as the initial tokenization) \cite{chen2015parallel}, fMLLR-based speaker adaptation \cite{heck2017feature}, and use of HMM to capture temporal dependency \cite{ansari2017unsupervised}.

\subsection{Optimizing DPGMM clustering}
% {\color{red}[1. post-processing towards clustering results; 2. optimizing input features to clustering]}

DPGMM clustering has been shown to be a preferred method of frame labeling for unsupervised subword modeling \cite{chen2015parallel,heck2017feature}.
Nevertheless, one shortcoming of DPGMM is that it tends to produce over-fragmented speech units  \cite{heck2018dirichlet,wu2018optimizing}.
Different approaches have been proposed to tackle this problem. In \cite{heck2018dirichlet}, DPGMMs were replaced by the Dirichlet process mixture of mixtures model (DPMoMM) to enable multi-modal cluster inference.
% \cite{heck2018dirichlet}.
In \cite{wu2018optimizing}, small-sized clusters were merged based on low functional load \cite{martinet1970economie,hockett1955manual}. 
% These approaches are effective in controlling the size of the inferred clusters without information loss of the target language.
% , hence achieve  improvements in unsupervised subword modeling. 
In our work, this problem is tackled by a label filtering algorithm.
% which filters
% The similar problem of DPGMM clustering was also observed in our work. To tackle this,  a label filtering algorithm is proposed to filter 
% out infrequently occurred DPGMM cluster labels. 

DPGMM for frame labeling could be optimized at input feature level. Conventional spectral features like MFCC \cite{chen2015parallel} and perceptual linear prediction (PLP) \cite{Heck+2016} were commonly used as the initial representations of target speech. Albeit straightforward, these features are considered sub-optimal for unsupervised subword modeling, as they contain a lot of irrelevant information such as speaker identity and emotion. Heck et al. \cite{heck2016unsupervised,Heck+2016} found that fMLLR transforms can noticeably suppress speaker-related feature variation, and advocated the importance of speaker adaptation in the concerned task. To enable supervised estimation of fMLLRs, clustering results on spectral features were taken as pseudo transcriptions. Chen et al. \cite{chen2017multilingual} showed that vocal tract length normalization (VTLN) on top of spectral features contribute to generating more robust DPGMM frame labels. 
% Pellegrini et al. \cite{pellegrini2017technical} adopted zero component analysis (ZCA) transformed features. 
In our study, fMLLR features are estimated by exploiting an out-of-domain ASR system.

\section{Proposed System}
\label{sec:framework}
\begin{figure*}[t]
\centering
\includegraphics[width=0.9 \linewidth]{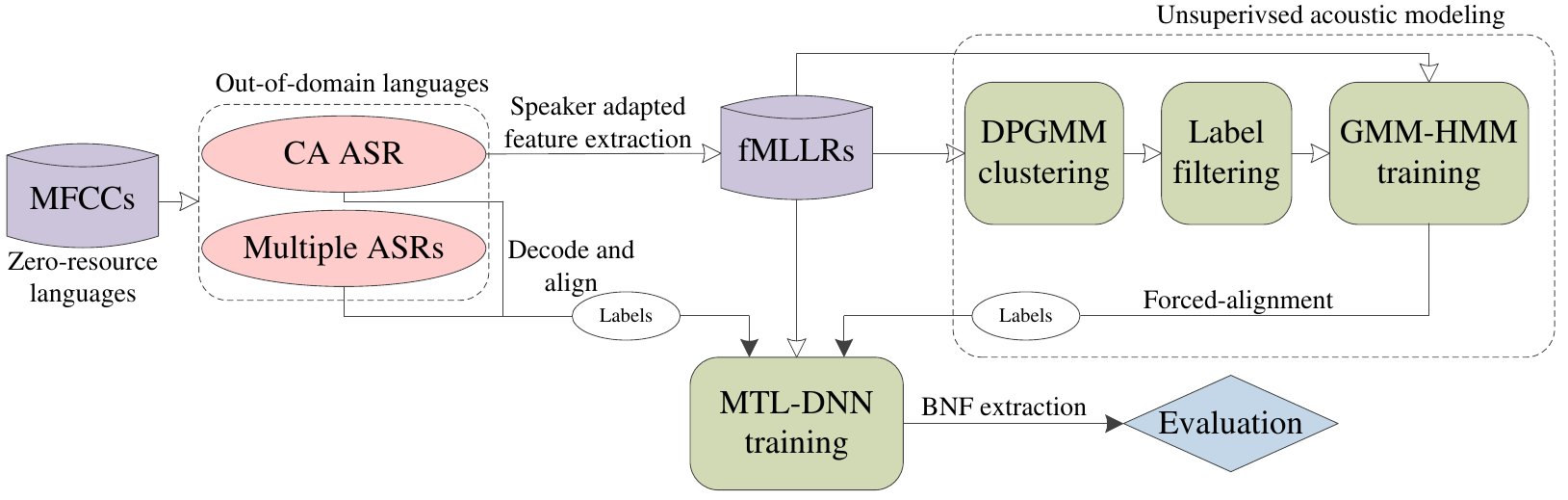}
\caption{The proposed framework of unsupervised subword modeling.}
\label{fig:framework}
\end{figure*}

%Fig. \ref{fig:framework} illustrates
The proposed system framework for unsupervised subword modeling of zero-resource languages is illustrated as in Fig. \ref{fig:framework}. It comprises three modules, namely, speaker-adapted feature extraction, unsupervised acoustic modeling, and multi-task BNF learning. Speech frames of the target language are first processed by an out-of-domain ASR system, where VTLN, LDA, MLLT and fMLLR transforms are estimated sequentially.
%speech frames are first assigned with labels, using either DPGMM clustering algorithm or out-of-domain ASR decoding.
The DPGMM clustering algorithm is applied to the fMLLR features of target speech. The resulted frame labels are post-processed by a label filtering algorithm and then used for context-dependent GMM-HMM (CD-GMM-HMM) acoustic modeling. The trained AMs forced align target speech to generate DPGMM-HMM alignments.
% with fMLLR-based speaker adaptive training (SAT).
 %After performing forced-alignment towards target speech by CD-GMM-HMM AMs, 
 Subsequently, an MTL-DNN is trained to generate BNFs for subword modeling. The training tasks of MTL include DPGMM-HMM  alignment prediction and language-mismatched label prediction  of multiple target languages. The language-mismatched labels are generated by multiple out-of-domain ASR systems.
 
%  of CD-GMM-HMM AMs of multiple target languages, and label prediction of multiple out-of-domain ASR systems.
% The out-of-domain ASR decodes target speech and provide an additional frame label to support multi-task learning DNN (MTL-DNN) training.
% We refer to our proposed framework as \emph{hybrid DPGMM-HMM-DNN}.

The proposed system design emphasizes on leveraging speech data resources from out-of-domain languages \cite{shibata2017composite,feng2018exploiting}. This is realized in the following aspects:
\begin{itemize}
\item Use out-of-domain data to perform fMLLR speaker adaptation on target speech.
\item Use out-of-domain ASR systems to generate frame labels to facilitate multi-task DNN training.
\item Use an out-of-domain DNN AM to extract BNFs.
%, with the goal of exploiting phonetic diversity in the mismatched language for target zero-resource subword modeling.

\end{itemize}

\subsection{Speaker adaptation with out-of-domain data}
\label{sec:spk_adapt}
For resource-rich languages, a large amount of transcribed and speaker-annotated speech data are readily available. We propose to utilize these out-of-domain data to model speaker variation in untranscribed speech of the target speech. A conventional CD-GMM-HMM AM is trained using the out-of-domain data. Based on this model, VTLN, LDA, MLLT and fMLLR transforms can be estimated. Subsequently, CD-GMM-HMM AMs with speaker adaptive training (CD-GMM-HMM-SAT) are trained and used to estimate fMLLR transforms for target speech utterances. It must be noted that the estimated fMLLR features of target speech could be directly used for subword modeling. They are expected to provide a better baseline than the conventional spectral features like MFCCs or PLPs.

\subsection{Frame labeling}
% Frame labeling is an essential step to prepare the target speech utterances for DNN-based subword discriminative modeling. While in some of the DNN models like AEs, frame labels are not needed, there were studies suggesting that AEs might not be a good approach to improving acoustic features in the zero-resource scenario \cite{renshaw2015comparison}.
% In the present study, two frame labeling approaches are investigated, namely, DPGMM clustering and out-of-domain ASR decoding.
%following works in \cite{chen2015parallel,heck2017feature,feng2018exploiting}, DPGMM is adopted for frame labeling.
\subsubsection{DPGMM clustering}
DPGMM is a non-parametric Bayesian extension to GMM, where a Dirichlet process prior replaces the vanilla GMM. One advantage of DPGMM clustering is that the cluster number does not need to be pre-defined.
Let us consider $M$ zero-resource target languages. For an utterance from the $i$-th language, the frame-level features are denoted as $\{\bm{x_1^{i}, x_2^{i}, \ldots, x_L^{i}}\}$, where $L$ is the number of frames in the utterance. By applying DPGMM clustering, $K$ clusters are obtained and represented with $k$ Gaussian components. The frame labels $\{l_1^{i}, l_2^{i}, \ldots, l_L^{i}\}$ are given as,
\begin{equation}
\label{eqt:dpgmm_inference}
l^{i}_{t} = \argmax_{1 \le k \le K} \mathrm{Prob}(k|\bm{x_t^i}),
\end{equation}
where $\mathrm{Prob}(k | \bm{x_{t}^{i}})$ denotes the posterior probability of $\bm{x_{t}^{i}}$ with respect to the $k$-th Gaussian component.
The inference of DPGMM parameters can be performed using the algorithm as described in \cite{chang2013parallel}.
% a Metropolis-Hastings based split/merge sampler was adopted \cite{chang2013parallel}. [**If you are using the same inference technique, you may say "we use the same algorithm as in chang2013 ... **]

\subsubsection{Out-of-domain ASR decoding}
%Frame labeling can also be done with an out-of-domain ASR system, which is typically trained with a large amount of transcribed speech in a resource-rich language.
Given a speech utterance in the target language, an out-of-domain ASR system can be applied to generate a sequence of phone-level or state-level labels \cite{feng2018exploiting}. The idea can be naturally extended to using multiple out-of-domain ASR systems and desirably providing a wide coverage of phonetic diversity. The outcome of ASR decoding depends on the relative weighting of AM and LM. In our work, the LM is assigned a very small weight, such that the acquired frame labels mainly reflect acoustic properties of the target speech being modeled.
%Input features for DPGMM clustering in this work are fMLLRs as estimated by an out-of-domain ASR system. This is inspired by the work in \cite{heck2017feature} that the speaker adapted features are shown to outperform raw spectral features for DPGMM clustering based unsupervised subword modeling. On the other hand, our previous work demonstrated the effectiveness and high efficiency of exploiting an out-of-domain ASR system to perform speaker adaptation towards target zero-resource speech \cite{feng2018exploiting}. This drives us to perform frame clustering based on fMLLRs instead of raw spectral features such as MFCCs.

\subsection{DPGMM label filtering}
\label{subsec:dpgmm_label_filtering}
\begin{figure}[t]
    \centering
    \includegraphics[width = \linewidth]{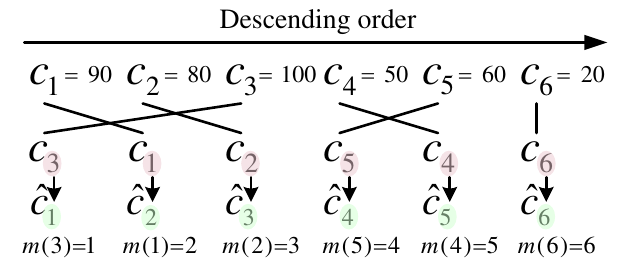}
    \caption{Example of cluster size sorting.}
    \label{fig:m_function}
\end{figure}

%The proposed approach processes frame labels of a certain language every time.
%Label filtering is performed for each zero-resource language separately. 
For a specific target language, let us assume that $K$ Gaussian components (clusters) are obtained by DPGMM clustering. The frame labels are denoted as $l_1, l_2, \ldots,l_N$ for an $N$-frame utterance.
%\begin{equation}
% \mathcal{S}=\Big\{l_1, l_2, \ldots, l_N\Big\}, l_i \in \Big\{1,2,\ldots, K\Big\}.
%\end{equation}
%where each label $l_i \in \{1,2,\ldots, K\}$.
Let $c_k$ be the number of frames labeled as cluster $k$, i.e.,
\begin{equation}
\label{eqt:c_k}
c_k = \sum_{i=1}^{N} \mathbb{1}(l_i = k), k \in \{1,2,\ldots,K\},
\end{equation}
where $\mathbb{1}(\cdot)$ is the indicator function.
%\begin{equation}
%c_k =  \left\vert\Big\{i : l_i = k, l_i \in \{1,2,\ldots, K\} \Big\}\right\vert.
%\end{equation}

% {\color{blue}
The elements in $\{c_1, c_2, \ldots, c_K\}$ are sorted in descending order into  $\{\hat{c}_1, \hat{c}_2, \ldots, \hat{c}_K | \hat{c}_1 \geq \hat{c}_2 \geq \ldots \geq \hat{c}_K \}$. $m(\cdot)$ denotes the index mapping function, i.e.,
\begin{equation}
\label{eqt:hat_c_k}
\hat{c}_k = c_{m(k)}.
\end{equation}
% {\color{blue} 
Fig. \ref{fig:m_function} gives an example of cluster size sorting.
% }

Let $P$ be the percentage of frame labels that we aim to retain. These frames are from $K_{cut}$ ``dominant'' clusters, where
\begin{equation}
\label{eqt:k_cut}
K_{\mathrm{cut}} = \argmin_{K^{\prime}}  \frac{\sum_{k=1}^{K^{\prime}} \hat{c}_k}{N} \geq P.
\end{equation}

$\mathcal{O}$ denotes the collection of all frame labels that are removed, i.e.,
\begin{align}
\label{eqt:o}
\mathcal{O}&=\Big\{l_i : l_i \in \mathcal{F}, i \in \{1,2,\ldots, N\} \Big\}, 
\end{align}
where
\begin{align}
\label{eqt:f}
\mathcal{F}&=\Big\{  m(K_{\mathrm{cut}}+1),\ldots, m(K)  \Big\}.
\end{align}

%cluster indices inside
$\mathcal{F}$ contains indices of $K-K_{cut}$ clusters that are the least frequent to occur. Frames assigned to these clusters are considered as outliers.

In the extreme case when $P$ is set to $1$, $\mathcal{F}$ and $\mathcal{O}$ will be empty sets. The smaller the value of $P$, the  larger the proportion of filtered frame labels. The label filtering algorithm is summarized as in Algorithm \ref{alg:dpgmm_label_filtering}.
\begin{algorithm}
\caption{DPGMM label filtering algorithm}\label{alg:dpgmm_label_filtering}
\hspace*{\algorithmicindent} \textbf{Input:} $l_1, l_2, \ldots,l_N$, $P$\\
\hspace*{\algorithmicindent} \textbf{Output:} $\mathcal{O}$
\begin{algorithmic}[1]
\State Calculate $c_k$ by Equation (\ref{eqt:c_k}).
\State Sort $\{c_1, c_2, \ldots, c_K\}$ in descending order. \label{alg:sort}
\State Calculate $m(k)$ by Equation (\ref{eqt:hat_c_k}).
\State Calculate $K_{\mathrm{cut}}$ by Equation (\ref{eqt:k_cut}) and $P$.
\State Select a subset of $l_1, l_2, \ldots,l_N$ as $\mathcal{O}$, by Equation (\ref{eqt:o})\&(\ref{eqt:f}).\Comment{Frame labels that are removed.}
%\Procedure{Euclid}{$a,b$}\Comment{The g.c.d. of a and b}
%\State $r\gets a\bmod b$
%\While{$r\not=0$}\Comment{We have the answer if r is c0}
%\State $a\gets b$
%\State $b\gets r$
%\State $r\gets a\bmod b$
%\EndWhile\label{euclidendwhile}
%\State \textbf{return} $b$\Comment{The gcd is b}
%\EndProcedure
\end{algorithmic}
\end{algorithm}
\subsection{DPGMM-HMM acoustic modeling}
\label{sec:ali}

Each DPGMM cluster can be regarded as a pseudo phone.
The sequence of DPGMM frame labels (after filtering) can be converted into a pseudo transcription by collapsing neighboring duplicated labels, e.g., \quotes{1,3,3,3,7,10,10} $\rightarrow$ \quotes{1,3,7,10}.  
% It is expected that contextual dependencies would be better modeled by HMM \cite{heck2017feature}. 
Based on the pseudo transcription, HMM acoustic modeling is done by following the standard supervised training pipeline, i.e., proceeding from monophone model training with uniform time alignment to context-dependent GMM-HMM (CD-GMM-HMM). 
% Note that in CI-GMM-HMM AM training,  
% The input features are fMLLR features estimated by an out-of-domain ASR, as discussed in Section \ref{sec:spk_adapt}. At the stage of CD-GMM-HMM training, feature transforms including LDA, MLLT and fMLLR are estimated based on the input fMLLR features.
% {\color{cyan}[** Are these input features fMLLR themselves ?**yes]}. 
The trained AM is used to produce time alignment information for DNN-based subword discriminative modeling (will be discussed in Section \ref{sec:mtl_bnf}). To be distinguished from the DPGMM frame labels, the frame labels obtained from the HMM forced alignment are referred to as \emph{DPGMM-HMM labels}.

Although the DPGMM labels could be directly used for supervised DNN acoustic modeling \cite{chen2017multilingual,feng2018exploiting}, we expect that DPGMM-HMM labels are more reasonable as they are derived with consideration on contextual dependency of speech.

%DPGMM has been widely applied in unsupervised speech modeling, especially in unsupervised word clustering \cite{kamper2014unsupervised},
%\begin{figure}[t]
%\includegraphics[width = 0.9\linewidth]{MTL_tnr_embed.pdf}
%\end{figure}

\subsection{Multi-task learning for BNFs}
\label{sec:mtl_bnf}
% [**Describe what is Bottleneck features **] 
The bottleneck feature (BNF) is a type of representation obtained from a designated low-dimension hidden layer of a DNN. 
In ASR applications, BNFs have been shown to provide a compact and phonetically-discriminative representation of input speech, and be effective in suppressing linguistically-irrelevant variations \cite{grezl2009investigation}. In the context of zero-resource speech modeling, BNFs have also been widely investigated \cite{chen2017multilingual,shibata2017composite,feng2018exploiting,hermann2018multilingual_journal}.
 %A recent work by \cite{hermann2018multilingual} demonstrated that multilingual BNFs perform .
% In this study, a multi-task learning (MTL) DNN \cite{caruana1998multitask} is adopted in order to leverage the phonetic diversity in different speech tasks and different languages.

\begin{figure}[t]
\centering
 \includegraphics[width =  \linewidth]{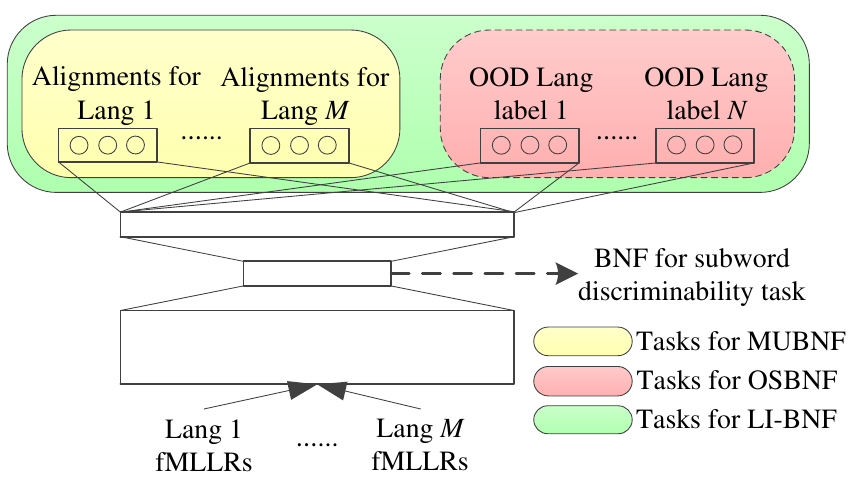}
\caption{MTL-DNN for extracting LI-BNF, MUBNF and OSBNF. The term ``OOD'' stands for out-of-domain.}
\label{fig:mtl}
\end{figure}

The proposed MTL-DNN is depicted  in Fig \ref{fig:mtl}. The DNN training involves a total of $M+N$ tasks, which involves
$M$ zero-resource target languages and $N$ out-of-domain ASR systems. Each of the tasks is represented by a task-specific softmax output layer in the DNN. The hidden layers, including a low-dimension linear BN layer, are shared across all tasks.
For the zero-resource language tasks, state-level or phone-level DPGMM-HMM labels
% [** Is it just DPGMM-HMM label you defined earlier ? If so, why not use the defined short name ?**] 
are used as target labels. The decoding output from each of the out-of-domain ASR systems  provides one set of frame-level labels for MTL.
%To make terminologies consistent with our previous work \cite{feng2018exploiting},
% For convenience, f

For the MTL-DNN trained only on the $M$ target language tasks, the extracted BNFs are referred to as multilingual unsupervised BNFs (MUBNFs). When out-of-domain ASR tasks are added, the BNFs are named language-independent BNFs (LI-BNFs). In the case that only the out-of-domain ASR tasks are involved, the extracted BNFs are referred to as out-of-domain supervised BNFs (OSBNFs). 

The DPGMM-HMM labels are obtained through statistical modeling of target speech. The ASR senone labels leverage the phonetic knowledge acquired from out-of-domain languages. It is expected that they would contribute complementarily in feature learning. Learning from speech of multiple languages would result in a language-independent BNF representation that is more generalizable to unknown languages.

For the shared-hidden-layer structure in the MTL-DNN, multi-layer perceptron (MLP) is commonly used \cite{chen2017multilingual,ansari2017deep,tsuchiya2018speaker,feng2018exploiting}. In this study, in addition to MLP, we investigate the use of long short-term memory (LSTM) \cite{sak2014long} and bi-directional LSTM (BLSTM) \cite{graves2013hybrid}, which were shown to perform better than MLP in conventional supervised acoustic modeling.

% {\color{blue}
On the other hand, BNF representation can also be obtained from the DNN AM pre-trained for a resource-rich language \cite{shibata2017composite}. This is considered as a transfer learning approach \cite{swietojanski2012unsupervised}. This transfer learning BNF (TLBNF) is expected to further enrich the feature representation and will be jointly used with MUBNF, OSBNF and LI-BNF for subword modeling.
% }
%The major  difference between TLBNFs  and MUBNFs, OSBNFs and LI-BNFs is that, BNF extractor for TLBNFs are trained using out-of-domain transcribed speech data, instead of in-domain zero-resource languages' data. 
%The generation of TLBNF representation can be considered as a form of transfer learning of linguistic knowledge from an out-of-domain resource-rich language to the target zero-resource languages, and is expected to carry complementary information to other forms of BNFs mentioned above.
% To boost the idea of transfer learning in unsupervised feature representation learning,
% In this work, two types of feature concatenation are investigated:
% \begin{itemize}
% \item[1.] Concatenating TLBNFs and LI-BNFs;
% \item[2.] Concatenating TLBNFs, MUBNFs and OSBNFs.
% \end{itemize}
% concatenating MUBNF, OSBNF and TLBNF

\section{Experimental setup}
\label{sec:exp}
\subsection{Dataset and evaluation metric}
% To evaluate our methods of unsupervised feature representation learning, and to facilitate direct comparison between our system and state-of-the-art works by other researchers, the
Experiments are carried out with the development data of ZeroSpeech 2017 Track 1 \cite{dunbar2017zero}.
% , \textit{unsupervised subword modeling} task \cite{dunbar2017zero}.
The data covers three target languages, namely English, French and Mandarin. For each language, there are separate training set and test set of untranscribed speech. Speaker identity information is provided for the train sets but not available for the test sets. The test data are organized into subsets of different utterance lengths: $1$ second, $10$ second and $120$ second. Detailed information about the dataset are given as in Table \ref{tab:zr17_data}.

\begin{table}[!h]
\renewcommand\arraystretch{1}
\centering
\caption{Development data of ZeroSpeech 2017 Track 1}
\label{tab:zr17_data}

%\resizebox{1 \linewidth}{!}{
\begin{tabular}{lcc|c}
% \hline
 \toprule[1pt]\midrule[0.3pt]
 & \multicolumn{2}{c|}{ Training} & Test \\
\midrule
 & Duration (hours) & \# speakers  & Duration (hours)\\
% \hline
\midrule
English & $45$  & $60$ & $27$ \\
French & $24$ & $18$ & $18$ \\
Mandarin & $2.5$  & $8$ & $25$ \\
% Training hours:  & $19.3$ & $81.5$ & $105.3$ \\
% Test hours:& $0.6$ & $0.7$ & $5.9$ \\
% Basic acoustic unit:  & Phone & Phone & Initial-Final \\
% \#basic units (inc. sil):& $33$ & $87$ & $61$ \\
% \#tied CD-HMM states:& $2462$ & $3431$ & $2386$ \\
% Lexicon size:& $ $ & $ 133K$& $ $ \\
% $\#$ Phonemes: &$43$& $46$&$ 29$& $44$& $38$\\
% \hline
\midrule[0.3pt]\bottomrule
\end{tabular}%
%}
% \label{tab:zr17_data}
\end{table}

The evaluation metric adopted for ZeroSpeech 2017 Track 1 task is the ABX subword discriminability. Inspired by the match-to-sample task in human psychophysics, it is a simple method to measure the discriminability between two categories of speech units \cite{versteegh2015zero}. The basic ABX task is to decide whether $X$ belongs to $x$ or $y$, if $A$ belongs to $x$ and $B$ belongs to $y$, where $A$, $B$ and $X$ are three data samples, $x$ and $y$ are the two pattern categories concerned. The performance evaluation in ZeroSpeech 2017 is carried out on the triphone minimal-pair task. A triphone minimal pair comprises two triphone sequences, which have different center phones and identical context phones, for examples, \quotes{beg}-\quotes{bag}, \quotes{api}-\quotes{ati}. Discriminating triphone minimal pairs is a non-trivial task. The performance of a feature representation on the triphone minimal-pair ABX task is considered a good indicator of its efficacy in speech modeling \cite{zeghidour2016deep}.
% Dynamic time warping (DTW) is used to measure the dissimilarity between each pair of speech segments, the underlying frame-level dissimilarity is allowed to be defined by participants. In this paper, cosine distance is selected as recommended by challenge organizers.

% {\color{cyan} [** see if the followings are OK ** modified statement in DTW and cosine distance]}
Let $x$ and $y$ denote a pair of triphone categories. Consider three speech segments $A$, $B$ and $X$, where $A$ and $X$ belong to category $x$ and $Y$ belongs to $y$. The ABX discriminability of $x$ from $y$ is measured in terms of the ABX error rate $\epsilon(x,y)$, which is defined as the probability that the distance of $A$ from $X$ is greater than that of $B$ from $X$, i.e.,
\begin{equation}
\begin{split}
 \epsilon(x,y) =& \frac{1}{\vert S(x) \vert (\vert S(x) \vert-1) \vert S(y) \vert} \sum_{A \in S(x)} \sum_{B \in S(y)}\sum_{X \in S(x) \backslash \{A\}} \\
 &(\mathbb{1}_{d(A,X) > d(B,X)} + \frac{1}{2} \mathbb{1}_{d(A,X) = d(B,X)}),
\end{split}
\end{equation}
where $S(x)$ and $S(y)$ denote the sets of features that represent triphone categories $x$ and $y$, respectively. $d(\cdot,\cdot)$ denotes the dissimilarity between two speech segments, which is computed by 
dynamic time warping (DTW) in our study. The frame-level dissimilarity measure used for DTW scoring is the cosine distance.
Note that $\epsilon(x,y)$ is asymmetric to $x$ and $y$. A symmetric form can be defined by taking average of $\epsilon(x,y)$ and $\epsilon(y,x)$. 
The overall ABX error rate is obtained by averaging over all triphone categories and speakers in the test set. A high ABX error rate means that the feature representation is not discriminative, and vice versa. Intuitively, the error rate should be no larger than $50 \%$, as by random decision, the expectation of ABX error rate is $50 \%$.

Two evaluation conditions were defined in ZeroSpeech 2017, namely \textit{within-speaker} and \textit{across-speaker}. In both conditions, the segments $A$ and $B$ to be evaluated are generated by the same speaker. In the within-speaker condition, segment $X$ is generated by the same speaker as $A$ and $B$; In the across-speaker condition, $X$ is generated by a speaker different from $A$ and $B$.
% This evaluation setting facilitates direct comparisons of speaker invariance in feature representation learning methods.
%, and enables the evaluation of these methods' generalization capabilities towards new speakers \cite{dunbar2017zero}.
% Depending on whether the segment $X$ belongs to the same speaker as $A$($B$),
%ABX error rates for \textit{within-speaker} and \textit{across-speaker} are evaluated separately, depending on whether $X$ and $A(B)$ belong to the same speaker.

\subsection{Out-of-domain ASR systems}
\label{subsec:ood_asr}
Four out-of-domain ASR systems are utilized and investigated in our experiments. They cover the languages of Cantonese (CA), Czech (CZ), Hungarian (HU) and Russian (RU).
%A Cantonese ASR is selected as the out-of-domain ASR system.
% {\color{cyan} [** The following description about training procedures needs to be re-organized. It should be made easier to understand by readers who may not be a skillful Kaldi user. **i added relation between CD-GMM-HMM-SAT and DNN-HMM.]}
The Cantonese ASR is trained with the CUSENT database \cite{LeeLoChingEtAl2002}. The database contains $20,378$ training utterances from $34$ male and $34$ female speakers, with a total of $19.3$ hours of speech. The Kaldi toolkit \cite{povey2011kaldi} is used to train two versions of AMs: CD-GMM-HMM-SAT and DNN-HMM. DNN-HMM training labels are acquired from CD-GMM-HMM-SAT time alignment. 
The input features for CD-GMM-HMM-SAT are $40$-dimension fMLLRs, and the input features for DNN-HMM are fMLLRs with $\pm 5$ splicing. The fMLLR features are estimated during CD-GMM-HMM-SAT training. Specifically, VTLN is estimated towards $39$-dimension MFCCs+$\Delta$+$\Delta\Delta$. The resulted features with $\pm 3$ splicing are used to estimate $40$-dimension LDA and MLLT. Finally, fMLLR transforms are estimated.
% They are estimated by  VTLN towards $39$-dimension MFCCs+$\Delta$+$\Delta\Delta$, and processed by splicing with context size $\pm 3$ to estimate $40$-dimension LDA and MLLT, followed by fMLLR estimation.
MFCC features are computed using a $25$-ms Hamming window and a $10$-ms frame shift. Per-utterance cepstral mean variance normalization (CMVN) is applied to MFCCs. 
% The total number of CD-HMM states is $2462$. 
The DNN-HMM model for Cantonese is a $7$-layer MLP, with layer configuration $440$-$1024\times 5$-$40$-$1024$-$2462$. The dimension of the output layer is determined by the number of CD-HMM states modeled by CD-GMM-HMM-SAT. Hidden layers are activated with sigmoid function, except for the $40$-dimension linear BN layer. The network is trained to optimize the cross-entropy criterion.
% . Sigmoid  is chosen as nonlinear activation function.
% The nonlinear function is Sigmoid, except for the linear bottleneck layer.
A syllable trigram LM trained with transcriptions of CUSENT training data is used during decoding. The LM is trained with SRILM  \cite{Stolcke02srilm--}.

The other three out-of-domain ASR systems are all phone recognizers developed by Brno University of Technology \cite{schwarz2009phoneme}. The recognizers adopt a $3$-layer MLP structure, in which the first two are sigmoid layers and the third is a softmax layer. They were trained with the SpeechDat-E databases \cite{heuvel2001speechdat}. The numbers of modeled phones in Czech, Hungarian and Russian are $45, 61$ and $52$, respectively. The training data sizes are $9.7$, $7.9$ and $14.0$ hours, respectively. The cross-entropy criterion was used for MLP training.
% Training criteria is cross-entropy.  
%9.72 7.86 14.02

\subsection{Speaker adaptation of target speech}
The Cantonese ASR system is used to perform fMLLR-based speaker adaptation of target speech on the $39$-dimension MFCC features in a two-pass procedure. In the first pass, input speech utterances are decoded in a speaker-independent manner, using unadapted features, from which initial fMLLR transforms are estimated. In the second pass, input speech are decoded with initial fMLLRs in a speaker-adaptive manner. After the decoding, final fMLLR transforms for target speech utterances are estimated. The dimension of fMLLR features is $40$.

\subsection{DPGMM frame clustering and label filtering}
\begin{figure}[!tbp]
\centering
\includegraphics[width  = 1.0 \linewidth]{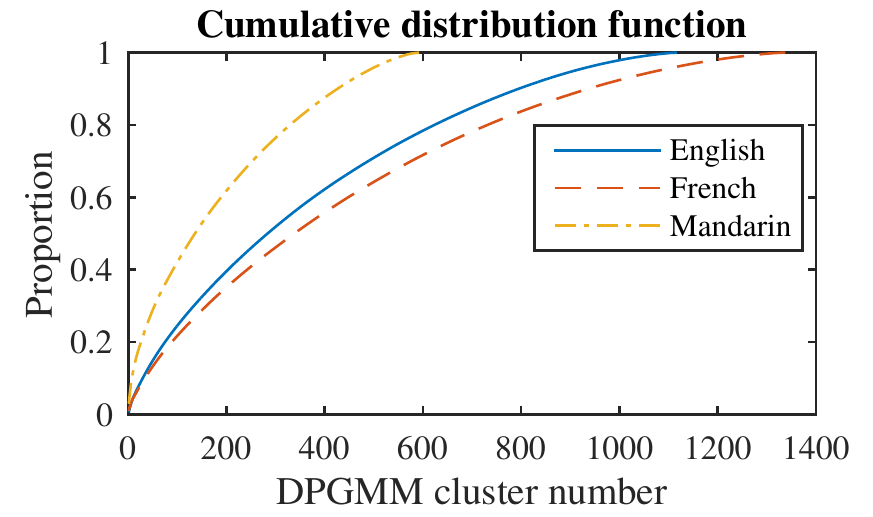}
\caption{Clustering results in the form of cumulative distribution function for the three target languages. Clusters are sorted according to cluster size in descending order.}
% \caption{Cumulative distribution functions of frame labels for the three target languages. DPGMM clusters are sorted according to cluster size in descending order.}
\label{fig:cdf}
\end{figure}

Speech frames for different languages are clustered separately by the DPGMM algorithm based on the $40$-dimension fMLLR features.
The implementation of DPGMM clustering is performed using an open-source tool developed by Chang et al. \cite{chang2013parallel}.
For the three target languages, namely English, French and Mandarin, the numbers of iterations of  clustering
were
% were empirically determined to be 
$120,200$ and $3000$ respectively.
% {\color{cyan} [** Are the following details really required ? And it reads to me that for all three languages, they are empirically determined **]}
% {\color{blue}
The numbers of iterations for English and French are determined by preliminary experiments. Specifically, the iterations for English ranging in $\{40, 80, \ldots , 680\}$ and for French ranging in $\{40, 80, \ldots, 400\}$ were tested. The optimal numbers of iterations were $120$ and $200$ respectively. For Mandarin, the number of iterations was empirically determined.
% } 
The resulted numbers of DPGMM clusters for English, French and Mandarin are $1118, 1345$ and $596$, respectively. Each frame is assigned a cluster label. 
Fig. \ref{fig:cdf} shows the results of clustering in the form of cumulative distribution function (CDF) for the three target languages. The clusters are sorted according to their cluster size in descending order. In other words, each point $(K_i, Q_i)$ on the CDF represents the  proportion of frame labels $Q_i$ that the largest $K_i$ clusters cover.

For label filtering, we evaluated different thresholds on the percentage of preserved labels, with the value of $P$ ranging from $0.6$ to $0.95$, with the step size of $0.05$.
% An illustration of cumulative distribution functions with respect to DPGMM clusters for English, French and Mandarin training sets is shown in Fig. \ref{fig:cdf}.
After filtering, the frame-level label sequences are converted into pseudo transcriptions, for the training of DPGMM-HMM AMs (in Section \ref{subsec:exp_hybrid_modeling}). 
%  collapsing consecutive repetitive labels, 
%  to form phoneme-like  pseudo {\color{blue}transcriptions} for downstream GMM-HMM acoustic modeling.
%  , as is done in conventional supervised GMM-HMM training pipeline.
 %It must be noted that while performing label filtering, we are not aiming at removing any speech frames from training sets, instead we focus only on frame label removal.

%In fact, the pseudo transcriptions needed for subsequent training are not frame-wise
%do not remove any target zero-resource speech frames from training sets, even if their corresponding labels are filtered out.
%FRAME FILTERING TBC
%\subsection{•}
%{\color{blue}[Feel unnatural]} 
% {\color{blue}\st{In addition to performing DPGMM clustering towards fMLLR features as mentioned above, we}}
%Note that the inputs to DPGMM clustering is not limited to fMLLR features. For comparison, we also 

DPGMM clustering was also tested with MFCCs as input features. The numbers of iterations for MFCC clustering are $200,240$ and $3000$ for English, French and Mandarin respectively, and the resulted numbers of DPGMM clusters are $1554,1541$ and $381$. 
% {\color{blue}\st{It serves as a baseline to fMLLR feature-based frame label acquisition procedure.}}

%to the clustering based on fMLLR features.

\subsection{DPGMM-HMM and MTL-DNN training }
\label{subsec:exp_hybrid_modeling}
%{\fontfamily{qcr}\selectfont
%This text uses a different font typeface
%}
% {\color{cyan} [** Here the name "hybrid DPGMM-HMM-DNN" appears the first time. It is not good to invent new name at this stage. I believe you can follow the terminologies of previous sections. **agree. this term is avoided.]}

DPGMM-HMM AMs are trained from scratch with pseudo transcriptions. Different from the conventional $3$-state HMM topology, during DPGMM-HMM training we set $1$-state HMM for each pseudo phone. This prevents the problem of unsuccessful forced alignments,
% unable to find alignments for speech utterances, 
as the numbers of pseudo phones for target languages are significantly larger than the number of phones for a typical language. The input features for DPGMM-HMM  are $40$-dimension fMLLRs estimated by the Cantonese ASR.  The  training procedure follows the standard pipeline as in Kaldi \textit{s5} recipe\footnote{\texttt{kaldi/egs/wsj/s5/run.sh}}, i.e., starting from CI-GMM-HMM to CD-GMM-HMM, followed by VTLN
%, LDA, MLLT estimation
and fMLLR-based SAT\footnote{LDA and MLLT are not estimated, as no improvement was found.}. After training, the numbers of CD-HMM states for English, French and Mandarin are $2818,2856$ and $2688$, respectively.

The MTL-DNN model is trained with all the three target zero-resource languages, from which  BNFs are extracted and evaluated by the ABX subword discriminability task. 
There are two types of tasks for MTL, namely, DPGMM-HMM alignment prediction task and out-of-domain ASR label prediction task.
%There are $4$ equally weighted tasks\footnote{We found through preliminary experiments that task weight tuning does not lead to performance improvement.}. 
In the first case, three tasks are included, i.e., frame alignments generated by DPGMM-HMM AMs, one for each target zero-resource language. 
In the second case, four tasks corresponding to Cantonese, Czech, Hungarian and Russian recognizers' senone labels are included. The senone labels are generated by decoding with LM to AM weight ratio set to $0.001$.
% During decoding, the LM to AM weight ratio is set to $0.001$.
% }}
%The last task is trained with alignments generated by the Cantonese ASR system, covering all three target languages. 
% {\color{blue}\st{Both CD-HMM state-level and phone-level alignments are tested in the training of MTL-DNN for comparison.}}
After MTL-DNN training, $40$-dimension HMM-LI-BNFs\footnote{The prefix `HMM-' emphasizes the use of DPGMM-HMM alignments, rather than DPGMM cluster labels.}  are extracted for the ABX task evaluation. Similarly, HMM-MUBNFs\footnotemark[3], extracted by MTL-DNN with DPGMM-HMM alignment tasks, and OSBNFs, extracted by MTL-DNN with one or more out-of-domain phone recognizers' senone labels, are also evaluated by the ABX task. The dimensions of both HMM-MUBNFs and OSBNFs are $40$. As illustrated in Fig. \ref{fig:mtl}, we defined several BNF representations according to the tasks included in MTL-DNN training. The configurations for (HMM-)MUBNF, OSBNF and (HMM-)LI-BNF are listed in Table \ref{tab:bnf_attrib}.

% {\color{cyan} [**The short name of Cantonese ASR is CA or CU ? Actually I think you can use the full names of the languages in the text. For the table of performance, you can use short names to save space.**] Thanks for reminder. I plan to use full names in text.}

\begin{table}[!h]
\renewcommand\arraystretch{1}
\centering
% {\color{blue}
\caption{Configurations for (HMM-)MUBNF, OSBNF and (HMM-)LI-BNF representations}
\label{tab:bnf_attrib}

\resizebox{ \linewidth}{!}{%
\begin{tabular}{l|ccc|ccc|cccc}
% \hline
 \toprule[1pt]\midrule[0.3pt]
Task label from& \multicolumn{3}{c|}{DPGMM}& \multicolumn{3}{c|}{DPGMM-HMM}& CA & CZ&HU&RU \\
\midrule
Train set & EN & FR & MA &  EN & FR & MA &  \multicolumn{4}{|c}{Pooling EN, FR and MA} \\
\midrule
MUBNF & \checkmark & \checkmark & \checkmark & & & & &&& \\
\midrule
OSBNF1 &  &  & &&&&\checkmark & & & \\
\midrule 
OSBNF2 & &&&&&& \checkmark & \checkmark & \checkmark & \checkmark \\
\midrule
LI-BNF1 &\checkmark & \checkmark & \checkmark &&&&\checkmark & & & \\
\midrule
LI-BNF2 &\checkmark & \checkmark & \checkmark&&& &\checkmark &\checkmark & \checkmark & \checkmark \\
\midrule
HMM-MUBNF & &&& \checkmark & \checkmark & \checkmark &&&&\\
\midrule
HMM-LI-BNF1 & &&& \checkmark & \checkmark & \checkmark & \checkmark &&&\\
\midrule
HMM-LI-BNF2 & &&& \checkmark & \checkmark & \checkmark  &\checkmark &\checkmark & \checkmark & \checkmark\\
\midrule[0.3pt]\bottomrule
\end{tabular}%
}
% }
% \label{tab:bnf_attrib}
\end{table}

% {\color{cyan} [** Is the following sentence repeating ? **agree]\st{Note that decoding results generated by an out-of-domain ASR depend on the relative weighting of AM and LM. In this work the LM to AM weight ratio is set to $0.001$, to ensure the acquired frame alignments mainly reflect acoustic properties of target speech.}}

The MTL-DNN is implemented in three different model structures: MLP, LSTM and BLSTM. The input features are $40$-dimension fMLLRs spliced with context size $\pm 5$. The dimensions of shared hidden layers in the MLP are $440$-$1024 \times 5$-$40$-$1024$. Sigmoid activation is used in all hidden layers except that the $40$ neurons in the BN layer use linear activation functions. The learning rate for MLP training is set at $0.008$ at the beginning, and halved when no improvement is observed on a cross-validation set. The mini-batch size is $256$. The LSTM model comprises $2$ LSTM layers with $320$-dimension cell activation vectors, and $1024$-dimension  outputs.
% {\color{cyan} [** what is ``layer output'' ? ** output dimension]}. 
A $40$-dimension BN layer followed by a $1024$-dimension fully-connected (FC) layer is set on top of LSTMs. For the BLSTM model, there are $2$ pairs of forward and backward LSTM layers. Each bi-directional layer has $320$-dimension cell activation vectors and $512$-dimension outputs. A BN layer followed by an FC layer is set on top of BLSTMs, with the same configuration as in the LSTM. The activation function in (B)LSTMs is $\tanh$. The learning rate is $2\mathrm{e}{-4}$ initially, and halved under the same criteria as for MLP. The truncated back-propagation through time (BPTT) algorithm \cite{williams1990efficient} is used to train (B)LSTM, with a fixed time step $T_{bptt}=20$. Note that the model parameters of LSTM and BLSTM structures were tuned in preliminary studies, while for MLP we follow the configuration of our previous study \cite{feng2018exploiting}.
% {\color{cyan}[Since feature concatenation subsection discarded, we have to report how TLBNF is generated from somewhere else.]}

\subsection{TLBNF generation}
The TLBNFs  for target zero-resource languages are generated by  applying the Cantonese DNN-HMM AM as the feature extractor. 
During TLBNF extraction, all the parameters of the DNN-HMM are fixed. 
% When  parameters of the DNN-HMM are frozen, 
The fMLLR features for target languages are fed as inputs to the DNN-HMM till its BN layer to generate TLBNFs.

\section{Results and Discussion}
\label{sec:results_and_analyses}
%%%%%%%%%%%%%%%%%%%%%%%%%
\begin{table*}[t]
\renewcommand\arraystretch{1}
\centering
\caption{ABX error rates ($\%$) on the baseline, our proposed methods and state of the art of ZeroSpeech 2017. MLP is adopted as the shared-hidden-layer structure. Label filtering is not applied.}
\label{tab:results_main}

\resizebox{1 \linewidth}{!}{%
\begin{tabular}{clccccccccc|c||ccccccccc|c}      
% \hline      
 \toprule[1pt]\midrule[0.3pt]
 & & \multicolumn{10}{c||}{ Within-speaker} & \multicolumn{10}{c}{ Across-speaker} \\
\midrule
 & & \multicolumn{3}{c}{ English} & \multicolumn{3}{c}{ French} & \multicolumn{3}{c|}{Mandarin}& Avg.&\multicolumn{3}{c}{ English} & \multicolumn{3}{c}{ French} & \multicolumn{3}{c|}{Mandarin} & Avg.\\
 && 1s & 10s & 120s & 1s & 10s & 120s & 1s & 10s & 120s && 1s & 10s & 120s & 1s & 10s & 120s & 1s & 10s & 120s &\\ 
% \midrule
%  & Duration & \#speakers  & Duration\\
% \hline
\midrule
\multirow{ 3}{*}{\circled{1}} & MFCC Baseline \cite{dunbar2017zero}& $12.0$ & $12.1$ & $12.1$ & $12.5$ &$12.6$ &$12.6$ &$11.5$ &$11.5$ & $11.5$&$12.0$ & $23.4$& $23.4$& $23.4$& $25.2$&$25.5$ &$25.2$ &$21.3$ &$21.3$ &$21.3$ & $23.3$ \\

& Out-of-domain fMLLR  \cite{feng2018exploiting} & $8.0$ & $8.2$ &$ 7.3$ & $10.3$ & $10.3$ & $9.1$ & $9.3$ &$ 9.3$ &$ 8.4$ & $8.9$ & $13.4$ & $12.0$ & $11.3$ & $17.2$ & $15.8$ & $14.8$ &$ 10.7$ & $10.2$ & $9.4$ & $12.8$  \\
& Out-of-domain fMLLR  \cite{shibata2017composite} & $7.8$  &$7.7$ &$7.0$ &$10.4$ &$10.5$ &$9.2$ &$9.2$ &$11.4$ &$8.8$ &$9.1$ &$14.2$ &$11.9$ &$11.3$ &$17.6$ &$15.2$ &$14.4$ &$12.7$ &$13.6$ &$10.0$ & $13.4$ \\
%{Heck et al.\color{blue}: fMLLR \cite{heck2017feature} } & $6.9$ & $6.4$ &$6.1$  & $10.3$ & $9.0$ &$8.6$  & $9.0$  &$8.2$  & $7.8$ &$8.0$ &$10.0$& $9.2$&$8.8$ &$13.9$ &$12.2$ &$11.8$ &$9.2$ &$7.7$ &$7.5$ & $10.0$ \\
\midrule
\multirow{ 6}{*}{\circled{2}}&MUBNF0 & $8.0$&$7.3$ &$7.3$&$10.3$ &$9.4$ &$9.3$ &$10.1$ &$8.8$ &$8.9$ & $8.8$ & $13.5$ & $12.4$&$12.4$ & $17.8$&$16.4$ &$16.1$ &$12.6$ &$11.9$ &$12.0$ &$13.9$\\
&MUBNF & $7.4$ & $6.9$ & $6.3$ & $9.6$ & $9.0$ & $8.1$ & $9.8$ & $8.8$ & $8.1$ & $8.2$ & $10.9$ & $9.5$ & $8.9$ & $15.2$ & $13.0$ & $12.0$ & $10.5$ & $8.9$ & $8.2$ & $10.8$ \\

&OSBNF1 & $7.2$ 	&$7.1$ &$6.3$ 	&$10.2$& 	$9.7$& 	$8.7$& 	$9.1$& 	$8.6$& 	$7.6$& 	$8.3$ & $10.0$& 	$9.7$& 	$8.6$& 	$13.9$& 	$13.4$& 	$11.6$& 	$9.0$& 	$8.4$& 	$7.5$& 	$10.2$ 
 \\
&OSBNF2 & $6.8$& 	$6.7$& 	$5.9$& 	$9.5$& 	$9.2$& 	$8.3$& 	$9.7$& 	$8.9$& 	$8.0$& 	$8.1$& $9.5$& 	$9.2$& 	$7.9$& 	$13.1$& 	$13.0$& 	$11.3$& 	$9.4$& 	$8.7$& 	$7.9$& 	$10.0$ \\ 
&LI-BNF1 & $6.9$& 	$6.6$& 	$6.1$& 	$9.5$& 	$9.2$& 	$8.4$& 	$9.2$& 	$8.5$& 	$7.9$& 	$8.0$&  
$10.0$& 	$8.9$& 	$8.2$& 	$14.3$& 	$12.9$& 	$11.5$& 	$9.5$& 	$8.5$& 	$7.7$& 	$10.2$ \\

&LI-BNF2 & $6.6 $&	$6.4$& 	$5.7$& 	$9.1$& 	$9.3$& 	$8.2$& 	$9.5$& 	$8.7$& 	$8.1$& 	$8.0$& 
$9.4$& 	$8.7$& 	$7.8$& 	$13.4$& 	$12.7$& 	$11.0$& 	$9.3$& 	$8.6$& 	$7.7$& 	$9.8$ 
\\
\midrule
\multirow{ 4}{*}{\circled{3}}&HMM(S)-MUBNF & $7.2 $&$	6.7 $&$	6.3 $&$	9.7 $&$	9.2 	$&$8.3 	$&$10.4 	$&$9.2 $&$	8.5 	$&$8.4 $ & $10.2 $&$	9.3 $&$	8.6 	$&$14.5 $&$	13.0 $&$	11.9$&$ 	10.7 $&$	9.2 	$&$8.4 $&$	10.6 $\\
&HMM(P)-MUBNF & $ 7.1 $&$	6.6 	$&$6.2 $&$	9.4 	$&$9.1 	$&$7.8 $&$	9.9 $&$	8.8 	$&$8.2 $&$	8.1 $ & $ 10.4 $&$	9.2 $&$	8.7 	$&$14.5 $&$	12.7 $&$	11.7 $&$	10.4 	$&$8.9 	$&$8.2 $&$	10.5 $\\
&HMM(P)-LI-BNF1 & $6.8 $&$	6.3 $&$	5.8 	$&$9.1 $&$	8.7 	$&$7.8 $&$	9.1 $&$	8.5 $&$	7.6 	$&$7.7 $ & $9.7 $&$	8.7 $&$	8.0 $&$	13.7 $&$	12.3 	$&$11.1 	$&$9.7 $&$	8.4 $&$	7.6 $&$	9.9 $\\
&HMM(P)-LI-BNF2 & $6.6$&$ 	6.4$ 	&$5.7$ 	&$9.2$& 	$8.8$& 	$8.1$& 	$9.2$& 	$8.6$& 	$7.9$& 	$7.8$  & $9.3$& 	$8.7$& 	$7.8$& 	$13.0$& 	$12.4$& 	$11.0$& 	$9.5$& 	$8.5$& 	$7.7$& 	$9.8$ \\
\midrule
\multirow{ 7}{*}{\circled{4}}&TLBNF & $7.2 $&	$6.8$& 	$6.1$& 	$9.6$& 	$9.0$& $	8.0$& 	$8.7$& 	$7.6$& 	$6.8$& 	$7.8$&   $10.6$& 	$9.6$& 	$8.7$& 	$14.2$& 	$13.2$& 	$11.5$& 	$8.5$& 	$7.6$& 	$6.7$& 	$10.1$ \\
&TLBNF+LI-BNF1 & $7.0$ &$6.6$&$6.0$&$9.3$&$8.8$&$7.9$& $8.6$&$7.5$& $6.7$ & $7.6$ &  $10.3$ &$9.3$&$8.4$&$13.9$&$12.9$&$11.4$&$8.5$&$7.6$&$6.7$&$9.9$ \\
&TLBNF+LI-BNF2 & $7.1$& 	$6.6$& 	$6.0$& 	$9.4$& 	$8.9$& 	$7.8$& 	$8.7$& 	$7.5$& 	$6.8$& 	$7.6$ & $10.4$ 	&$9.4$ 	&$8.5$ 	&$14.0$ 	&$13.0$ 	&$11.3$ 	&$8.5$ 	&$7.6$ 	&$6.6$ 	&$9.9$
\\
&TLBNF+HMM(P)-LI-BNF1 & $7.0$& 	$6.6$& 	$6.0$& $9.4$& $8.8$& 	$7.8$& 	$8.6$& 	$7.5$& 	$6.7$& 	$7.6$ &  $10.3$&$9.4$&$	8.4$&$13.9$&$ 	12.9$&$ 	11.3$&$ 	8.5$&$ 	7.6$&$ 	6.6$&$ 	9.9$\\

%TLBNF+HMM(P)LI-BNF2 & \\
&TLBNF+MUBNF+OSBNF1 & $6.8$ &	$6.4$& 	$5.8$ &	$9.0$ &	$8.8$ &	$7.8$& 	$8.5$& 	$7.7$& 	$6.8$& 	$7.5$ & $9.9$ &	$9.0$ &	$8.2$ &	$13.6$& 	$12.6$ &	$11.1$ &	$8.4$ &	$7.7$ &	$6.7$ &	$9.7$\\
&TLBNF+HMM(P)-MUBNF+OSBNF1 & $6.8$ &	$6.4$ &	$5.7$ &	$8.8$ &	$8.7$ &	$7.5$ &	$8.4$ &	$7.5$ &	$6.8$ &	$7.4$& $10.0$ &	$9.0$ &	$8.2$ &	$13.6$ &	$12.6$ &	$11.1$ &	$8.4$ &	$7.6$ &	$6.7$ &	$9.7$\\

&TLBNF+HMM(P)-MUBNF+OSBNF2 & $6.7$& 	$6.4$& 	$5.8$& 	$9.0$& 	$8.8$& 	$7.5$& 	$8.3$& 	$7.5$& 	$6.8$& 	$7.4$&  $10.0$& 	$9.0$& 	$8.2$& 	$13.6$& 	$12.6$& 	$11.1$& 	$8.4$& 	$7.6$& 	$6.7$& 	$9.7$\\
\midrule

&Heck et al. \cite{heck2017feature}& $6.9$&$6.2$&$6.0$&$9.7$&$8.7$&$8.4$&$8.8$&$7.9$&$7.8$ &$7.8$ &$10.1$ &$8.7$&$8.5$&$13.6$&$11.7$&$11.3$&$8.8$&$7.4$&$7.3$&$9.7$\\
&Chorowski et al. \cite{chorowski2019unsupervised} & $5.8$ & $5.7$ & $5.8$ & $7.1$ & $7.0$ & $6.9$ & $7.4$ & $7.2$ & $7.1$ & $6.7$ & $9.3$ & $9.3$ & $9.3$ & $11.9$ & $11.4$ & $11.6$ & $8.6$ & $8.5$ & $8.5$ & $9.8$\\
\midrule[0.3pt]\bottomrule
\end{tabular}%
}
% \label{tab:results_main}
\end{table*}

Table \ref{tab:results_main} provides a master summary to facilitate performance comparison among different systems of feature representation learning. The methods are organized in four groups, marked by circled numerals \circled{1} to \circled{4}  in the Table. The first group comprises a few relevant baseline and reference systems. The MFCC baseline system refers to the one, in which generic MFCC features are directly used in triphone minimal pair discrimination. 
The first out-of-domain fMLLR system comes from  previous work \cite{feng2018exploiting}, which used a Cantonese ASR system for fMLLR estimation.
The second one used a Japanese ASR \cite{shibata2017composite}.

The second and third groups of systems all use multilingual BNF representations, which are learned by different methods as described in Section \ref{subsec:exp_hybrid_modeling}. DPGMM labels and DPGMM-HMM labels are applied in the the second group and the third group respectively. In the second group, MUBNF0 is learned using MFCC as input features for DPGMM clustering and MTL-DNN modeling.
% while MUBNF are trained using fMLLRs as inputs.
The other representations in these two groups are learned using fMLLRs as DNN input features. As described in Section \ref{subsec:exp_hybrid_modeling} and Table \ref{tab:bnf_attrib}, OSBNF1 and OSBNF2 are trained with out-of-domain ASR senone labels, and LI-BNF1 and LI-BNF2 are trained with both DPGMM labels and out-of-domain ASR senone labels. 
In the third group, ``HMM(S)'' and ``HMM(P)'' denote the use of state-level and phone-level HMM alignments respectively for label generation.
% HMM(P) denotes the use of phone-level time alignment for frame label generation in MTL-DNN training, while HMM(S) denotes the use of state-level alignment.
The fourth group of systems are built on different combination of BNF features. The ``+''  sign is used to denote concatenation of two frame-level feature representations. 
The experimental results on all methods of BNF representation learning as shown in Table \ref{tab:results_main} are obtained by using the MLP structure in MTL-DNN. 
% The comparison among various neural network structures will be discussed in {\color{cyan}?} 
In addition, two representative systems that achieved very good performances in ZeroSpeech 2017 \cite{heck2017feature,chorowski2019unsupervised} are also listed in the Table.

% , namely, Heck et al. \cite{heck2017feature} and Chorowski et al. \cite{chorowski2019unsupervised}.

% {\color{cyan}\st{ The sign "+" denotes feature concatenation. Unless specified, BNFs listed in this Table are trained with fMLLR features as inputs.
%It can be seen from this
% From this Table, several  observations can be made as listed below: }}
\subsection{Effect of out-of-domain speaker adaptation}
% Experimental results of our proposed methods without label filtering, official baseline and state of the art of ZeroSpeech 2017{\color{blue} \cite{heck2017feature,chorowski2019unsupervised}}
%official baseline and our proposed methods

% as the shared-hidden-layer structure for MTL-DNN in all cases.
%\footnote{Part of results were reported in our past work \cite{feng2018exploiting}.}.
% as labels (``P'' stands for phone-level, or state-level if not specified).
% Otherwise DPGMM clustering results, i.e., cluster indices assigned to frames, are used as labels.
%the DNN training labels are obtained by forced-alignments of DPGMM-HMM AMs, i.e., DPGMM-HMM labels. Otherwise the labels are DPGMM clustering results, i.e., DPGMM labels.
%For DGPMM-HMM labels, there are two levels of labels feasible for DNN training, namely, pseudo phone-level\footnote{Each DPGMM cluster is considered a phoneme-like acoustic subword unit.} alignments and CD-HMM state-level alignments. The term \quotes{Phn} denotes the former level, otherwise the latter level of labels is used.
%\quotes{Phn} denotes phone-level labels\footnote{We use the term \emph{phone-level} to represent that each DPGMM cluster denotes a phoneme-like acoustic subword unit.}, otherwise are CD-HMM state-level labels.
% Both state-level and phone-level DPGMM-HMM alignments are tested. In this Table, \quotes{Phn} denotes phone-level alignments are used for DNN training, otherwise state-level alignments are used.

%{\color{blue}Compare MUBNF trained with MFCC/fMLLR}
The fMLLR features estimated with in-domain data were shown to perform significantly better than conventional spectral features in unsupervised subword modeling \cite{Heck+2016,Heck2016iterative}. In the present study, it has been shown that similar improvement could also be attained by performing speaker adaptation using an out-of-domain ASR system. Both out-of-domain fMLLR features in the first group of systems outperform the MFCC baseline consistently on all target languages.
% , with relative ABX error rate reduction $25.8 \%$ in within-speaker and $45.1 \%$ in across-speaker conditions. 
This improvement can be achieved without requiring any transcribed training data of the target language, which is highly desirable in the zero-resource scenario.

% {\color{blue}
In \cite{shibata2017composite}, the out-of-domain ASR system was trained on $240$ hours of Japanese speech. The experimental results in \cite{feng2018exploiting} show that using a Cantonese ASR system trained on only $19$ hours of speech could give a better performance in both within- and across-speaker conditions. The advantage is particularly significant when the target language is Mandarin.
% }

% {\color{cyan}[*consider putting it to a later section*]It is noted that the fMLLRs consistently perform better on long utterances ($120$s) than on short ones ($1$s and $10$s).}
% This is in agreement with our expectation, as 
%can be explained by the commonly acknowledged fact that 
% does not work well on very short speech.
%, while 
% In contrast, baseline MFCCs perform uniformly on different utterance lengths. This is due to the fact that no utterance-level normalization was adopted for MFCC extraction.

\subsection{Effectiveness of multilingual BNFs}

The following observations can be made on the performances of the learned multilingual BNF representations:

(1) BNF representations learned by MTL-DNN clearly outperform the respective input features to the DNN. MTL-DNN training with DPGMM labels is effective for both MFCC  and fMLLR.
% MTL-DNN training with DPGMM labels significantly improves subword discriminability. 
% MUBNF0 and MUBNF both outperform their MTL-DNN input features to a large margin, by only employing purely data-driven frame labels.
The average ABX error rates achieved by MUBNF0 are $8.8\%$ and $13.9\%$ in the within-speaker and across-speaker conditions respectively, versus $12.0\%$ and $23.3\%$ attained by MFCC.
For MUBNF representation, the relative performance improvements over fMLLR are $7.9\%$ and $15.6\%$ in the two test conditions.
% These improvements are achieved by adopting purely data-driven DPGMM labels for subword discriminative modeling.
MUBNF outperforms MUBNF0 to a large extent, especially in the across-speaker test condition.
This suggests that speaker adaptation at input feature level is a  critical step in obtaining speaker-invariant BNF representations.

(2) The effectiveness of BNF can be further improved by training the MTL-DNN with additional out-of-domain ASRs' senone labels. 
With the Cantonese ASR's senone labels included as one of the training tasks, the LI-BNF1 representation reduces within-/across-speaker ABX error rates by absolute
%achieves additional relative error rate reduction of 
$0.2\%/0.6\% $ as compared to MUBNF.
%, which are trained with only $3$ DPGMM label prediction tasks. 
When the senone labels of Czech, Hungarian and Russian are added, the resulted LI-BNF2 representation shows a further improvement of absolute $0.4\%$ under the across-speaker condition. 
This shows that out-of-domain acoustic-phonetic knowledge provides complementary information to the in-domain clustering labels for feature learning.
% as the complementary information to in-domain information represented by DPGMM clustering results. 
The performance gain of OSBNF2 over OSBNF1, as well as that of LI-BNF2 over LI-BNF1, confirm the benefit of exploiting a wider coverage of language resources.

The performance of OSBNF2 is inferior to OSBNF1 on Mandarin test set, but not on English and French. It is noted that OSBNF1 is learned by using the Cantonese ASR senone labels while OSBNF2 is learned by involving Cantonese and the other three European languages. Cantonese, being a Chinese dialect, is apparently closer to Mandarin than Czech, Hungarian and Russian in terms of acoustic-phonetic properties. The experimental results imply that the frame labels generated by involving highly-mismatched out-of-domain languages may be of low quality and not suitable for feature learning.
%Mandarin and Cantonese both belong to Chinese languages, while English and French are considered largely different from Cantonese. 
%While training a DNN for OSBNF1, 
%the mismatch between source and target languages is believed to be weaker than that between  Cantonese and English/French.

(3)  
% {\color{blue}
% \st{DPGMM-HMM AM training is effective in improving BNF representation. In our system design, DPGMM-HMM AMs are trained to forced-align target speech and generate HMM state-/phone-level frame labels, so as to replace DPGMM clustering-based labels.}
As discussed in Section \ref{sec:ali}, DPGMM-HMM labels are obtained by modeling temporal dependency of speech and DPGMM labels are determined with the assumption that neighboring speech frames are independent. Comparing the corresponding systems in the second and the third groups of Table \ref{tab:results_main}, it is noted that DPGMM-HMM labels perform slightly better than DPGMM labels.
% } 
%Frame alignments generated by DPGMM-HMM AMs serve a better form of supervision for MTL-DNN based subword discriminative modeling, as compared to labels by DGPMM clustering results. 
%Moreover, phone-level frame alignments are better than state-level ones. 
The ABX error rates attained with HMM(P)-MUBNF, HMM(P)-LI-BNF1 and HMM(P)-LI-BNF2 are about absolute $0.2\%$ - $0.3\%$ lower than those with MUBNF, LI-BNF1 and LI-BNF2 respectively, except for HMM(P)-LI-BNF2 under the across-speaker  condition.
% , except that HMM(P)-LI-BNF2 achieves the same across-speaker ABX error rate as LI-BNF2.
% For LI-BNF representation, the advance of phone-level DPGMM-HMM labels over DPGMM labels is also observed. 
This demonstrates that capturing temporal dependency in speech is beneficial to feature learning for subword modeling \cite{Heck2016iterative}.
% frame alignments generated by DPGMM-HMM AMs serve a better form of target labels for DNN training, as compared to DPGMM  labels.
%In LSTM and BLSTM structures, the same observation is also made.
%This indicates the effectiveness of DPGMM-HMM acoustic modeling for improving subword discriminability. 
% In conparison to   DPGMM labels, 
% This is due to the fact that 
% DPGMM-HMM alignments successfully model the abundant  temporal dependencies in speech through the HMM structure. 
It is also noted that phone-level HMM alignments are better than state-level ones. 

(4) Combining different types of BNF feature representations leads to further improvement of performance. Specifically, by concatenating HMM(P)-MUBNF, OSBNF1 and TLBNF, the best ABX error rates under both within-speaker and across-speaker conditions are achieved ($7.4\%$ and $9.7\%$).
It is found that BNFs learned from in-domain unsupervised data (HMM(P)-MUBNF, OSBNF1) and learned via transfer learning (TLBNF) can be jointly used to compose an optimal feature representation that is better than any individual BNF.

The best performance attained in this study is competitive to the best submitted system for the ZeroSpeech 2017 challenge, which is based on the combination of multiple DPGMM posteriorgrams \cite{heck2017feature}. These posteriograms were generated with unsupervisedly estimated fMLLRs based on different implementation parameters. The combination of posteriorgrams led to $3.0\%$ and $3.3\%$ relative error rate reduction under the within-speaker and across-speaker conditions, compared to the use of single posteriorgram representation. In our work, concatenating the  three aforementioned BNF representations results in $5.1\%$ and $4.0\%$ relative error rate reduction, as compared with the best system with single BNF.
% the best representation among HMM(P)-MUBNF, OSBNF1 and TLBNF. {\color{cyan}[We cannot say `with single BNF' because the best single BNF is HMM(P)-LI-BNF1/2.]}
% with single BNF.  
% The system development 
It must be noted that no out-of-domain transcribed speech was involved in the system of \cite{heck2017feature}.

% The performance improvement via combination of various BNFs indicates the complementarity between BNFs generated from DNNs trained with target zero-resource speech copora, and those via transfer learning.

% {\color{blue}
% The comparison between fMLLR representations estimated in our study and by Heck et al. \cite{heck2017feature} (within-speaker: $8.0\%$; across-speaker: $10.0\%$) shows that speaker adaptation by an out-of-domain ASR system cannot perform as well as by exploiting unsupervised in-domain data.  
% This could be partially explained by the domain mismatch between in- and out-of-domain speech data, including language, recording condition, channel etc.
% } 

% {\color{blue}
In a very recent work \cite{chorowski2019unsupervised}, vector quantized VAE (VQ-VAE) was applied to develop a system of unsupervised subword modeling. The reported average ABX error rate was $6.7\%$ for within-speaker condition, which is the best among all reported systems so far. For the across-speaker condition, our proposed systems with combined BNF features have slightly better performance than VQ-VAE ($9.8\%$). Our systems are found to be more effective on long utterances than VQ-VAE. In Table \ref{tab:results_main}, it is noted that the performance of VQ-VAE does not depend on utterance duration. For English and Mandarin, the ABX error rates are almost exactly the same between the cases of $1$s and $120$s. One possible reason is that the VQ-VAE system does not perform explicit utterance-level speaker normalization on input features. On the contrary, the BNF representations investigated in the study perform significantly better on longer utterances ($10$s \& $120$s) than on $1$s ones. It is also noted that our systems are more effective for Mandarin in the across-speaker condition. This may be due to the use of Cantonese speech in feature learning. VQ-VAE may be over-fitting to Mandarin due to small data size \cite{chorowski2019unsupervised}.
% }

\subsection{Effectiveness of label filtering}
%%%%%%%%%%%%%%%
\begin{table*}[tbp]
\renewcommand\arraystretch{1.1 }
\centering
\caption{Comparison of MTL-DNN shared-hidden-layer structures in feature representation learning of ZeroSpeech 2017.}
\label{tab:results_dnn_structures}

\resizebox{1 \linewidth}{!}{%
\begin{tabular}{llccccccccc|c||ccccccccc|c}      
% \hline      
 \toprule[1pt]\midrule[0.3pt]
 \multicolumn{2}{c}{} & \multicolumn{10}{c||}{ Within-speaker} & \multicolumn{10}{c}{ Across-speaker} \\
\midrule
 \multicolumn{2}{c}{} & \multicolumn{3}{c}{ English} & \multicolumn{3}{c}{ French} & \multicolumn{3}{c|}{Mandarin}& Avg.&\multicolumn{3}{c}{ English} & \multicolumn{3}{c}{ French} & \multicolumn{3}{c|}{Mandarin} & Avg.\\
 \multicolumn{2}{c}{}& 1s & 10s & 120s & 1s & 10s & 120s & 1s & 10s & 120s && 1s & 10s & 120s & 1s & 10s & 120s & 1s & 10s & 120s &\\ 
% \midrule
%  & Duration & \#speakers  & Duration\\
% \hline
\midrule
\multirow{3}{*}{MUBNF}  & MLP& $ 7.4$ & $6.9$ &$ 6.3$ &$ 9.6$ &$ 9.0$ & $8.1$ &$ 9.8$ &$ 8.8$ &$ 8.1$ & $\bm{8.2} $&$ 10.9 $&$ 9.5 $&$ 8.9 $&$ 15.2 $&$ 13.0 $&$ 12.0 $&$ 10.5 $&$ 8.9 $&$ 8.2 $&$ \bm{10.8}$ \\
& LSTM & $ 7.4 $&$	7.1 $&$	6.8 $&$	10.0 $&$	9.5 	$&$8.7 $&$	10.4 $&$	9.5 	$&$8.7 $&$	8.7 $&$10.4 $&$	9.6 $&$	9.0 $&$	14.6 $&$	13.3 $&$	12.3 $&$	10.9 $&$	9.3 $&$	8.6 $&$	10.9 $\\
& BLSTM & $7.4$ &$	7.1$ &	$6.7 $	&$9.9 	$&$9.5$ 	&$8.9 $&	$10.4$ &	$9.4 $&	$8.7 $	&$8.7$ &$10.4$ &	$9.6 $	&$9.0$ &	$14.7$ &	$13.3$ &	$12.1$ &	$10.7 $&$	9.3 $	&$8.6 $	&$10.9$ \\
\midrule
\multirow{3}{*}{HMM(P)-MUBNF} & MLP & $ 7.1 $&$	6.6 	$&$6.2 $&$	9.4 	$&$9.1 	$&$7.8 $&$	9.9 $&$	8.8 	$&$8.2 $&$	\bm{8.1} $&$ 10.4 $&$	9.2 $&$	8.7 	$&$14.5 $&$	12.7 $&$	11.7 $&$	10.4 	$&$8.9 	$&$8.2 $&$	\bm{10.5} $\\
& LSTM & $7.2 $&$	6.8 $&$	6.4 $&$	9.9 $&$	9.4 	$&$8.7 	$&$10.4 	$&$9.5 $&$	8.8 $&$	8.6 $&$10.0 $&$	9.3$&$ 	8.6 $&$	14.3 $&$	13.1 $&$	11.8 	$&$10.7 $&$	9.3 	$&$8.6 	$&$10.6 $\\
& BSLTM  &$7.3$ &	$6.9$ &	$6.5$ &	$9.6$ &	$9.5 $&	$8.4$ &	$10.5$ &	$9.4$ &	$9.0$ &	$8.6$ &$10.1 $	&$9.4$ 	&$8.9$ &	$14.2$ 	&$13.0$ 	&$11.9$ &	$10.8$& 	$9.4$ 	&$8.7$ &	$10.7$\\
\midrule
\multirow{3}{*}{HMM(P)-LI-BNF1} & MLP &$6.8 $&$	6.3 $&$	5.8 	$&$9.1 $&$	8.7 	$&$7.8 $&$	9.1 $&$	8.5 $&$	7.6 	$&$\bm{7.7} $ &$9.7 $&$	8.7 $&$	8.0 $&$	13.7 $&$	12.3 	$&$11.1 	$&$9.7 $&$	8.4 $&$	7.6 $&$	\bm{9.9} $\\
& LSTM &$6.7 	$&$6.6 $&$	5.9 $&$	9.5 $&$	9.4 $&$	8.2 	$&$9.6 $&$	8.9 	$&$7.9 $&$	8.1 $ &$9.6$& 	$9.1 	$&$8.1 $	&$14.1$ 	&$13.3 $&	$11.6$ &	$10.2$ &$	9.1 	$&$8.0 $&	$10.3 $ \\
& BLSTM &$7.0$&$6.6$ 	&$6.1$ 	&$9.3$ 	&$9.2$ 	&$8.2$ 	&$9.4$ 	&$8.7$ 	&$8.0$ &$8.1$ & $9.5$ 	&$9.0$ 	&$8.2$ 	&$13.7$ 	&$13.0$ 	&$11.6$ 	&$9.7$ 	&$8.7$ 	&$7.8$ 	&$10.1$ \\
\midrule[0.3pt]\bottomrule
\end{tabular}%
}
% \label{tab:results_dnn_structures}
\end{table*}

\begin{figure}[!tbp]
\centering
\includegraphics[width = 1.0 \linewidth]{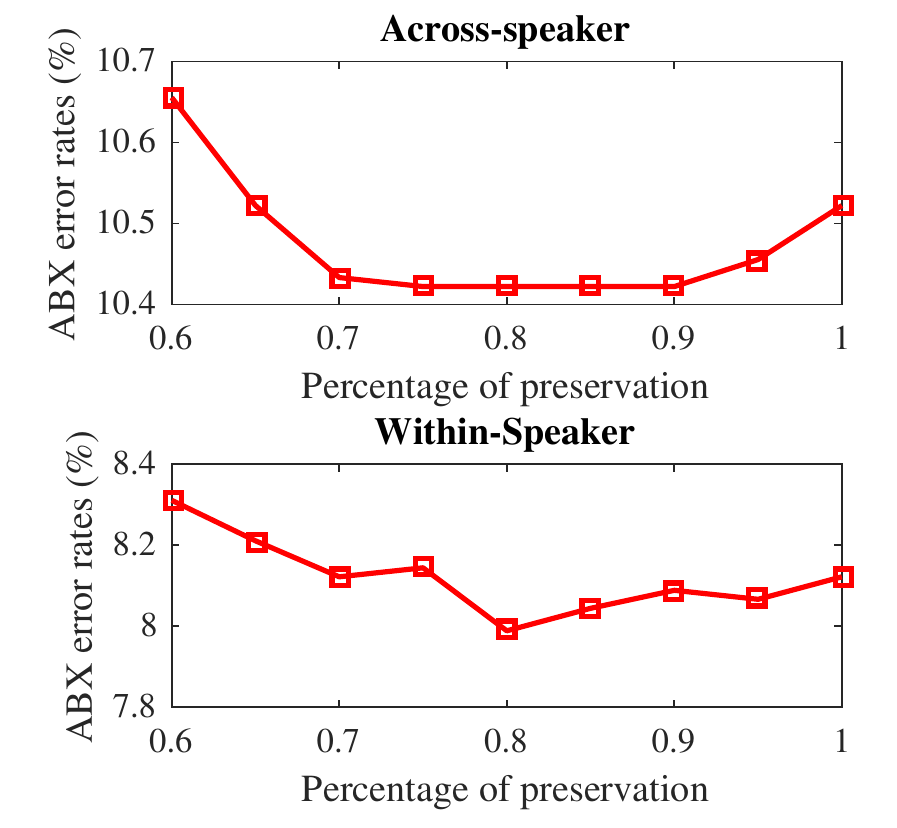}
\caption{Average ABX error rates ($\%$) with respect to label filtering percentage over three zero-resource languages, in HMM(P)-MUBNF representation.}
\label{fig:percent}
\end{figure}

The effectiveness of the proposed label filtering algorithm is evaluated with the HMM(P)-MUBNF representation, which is trained 
exclusively based on DPGMM-HMM labels, without involving out-of-domain speech data. Algorithm \ref{alg:dpgmm_label_filtering} requires one tunable parameter $P$, i.e., the percentage frame labels to be retained. The average ABX error rates attained with different values of $P$ are plotted as in Fig. \ref{fig:percent}.   $P=1$ means that all labels are kept, which is the setting used to obtain the results in Table \ref{tab:results_main}.

Under both within-speaker and across-speaker conditions, the optimal values of $P$ are in the range of $0.7$ to $0.9$. That is, when on average about $10-30\%$ of the frame labels are removed, the ABX error rates could be slightly reduced. This indicates that indeed a certain portion of the labels are not reliable. However, if too many labels are removed, e.g., more than $30\%$, the system performance would degrade significantly, because some good labels are lost.

% {\color{red}[pros/cons of our algo. like contextual information.][other lossless methods in \cite{heck2018dirichlet,wu2018optimizing}]}
% {\color{blue} 
The proposed label filtering method is very simple in that only the occurrence counts of the labels are considered. Fig. \ref{fig:percent} shows that this criterion is appropriate to a certain extent. However, there may exist infrequent subword units that are meaningful and crucial in conveying linguistic content. In \cite{heck2018dirichlet,wu2018optimizing}, it was suggested to reduce the number of DPGMM clusters without ignoring any frame labels. Since these studies were carried out on a different database, direct comparison of system performance can not be made.

\subsection{Comparison of DNN model structures}
\label{sec:compare_dnn_structures}

For BNF feature learning with the MTL-DNN approach, DNN models other than MLP can be used. Table \ref{tab:results_dnn_structures} compares the system performances obtained by using MLP, LSTM and BLSTM. The feature representations being investigated include MUBNF, HMM(P)-MUBNF and HMM(P)-LI-BNF1, and label filtering is not applied. 
% Note that hyper-parameters of LSTM and BLSTM are tuned by a set of preliminary experiments, while no tuning was performed for MLP. 
% and Fig. \ref{fig:comp_mlp_lstm_blstm}.

\begin{figure}[t]
\centering
\includegraphics[width =   \linewidth]{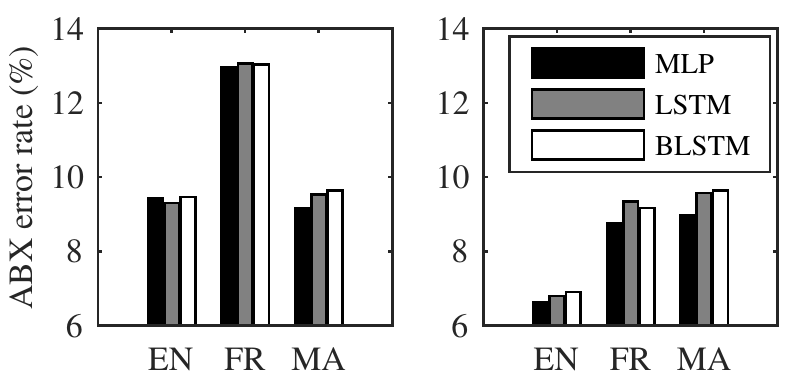}
\caption{Average ABX error rates ($\%$) of HMM(P)-MUBNF representation over utterance lengths for each language. Left: Across-speaker; Right: Within-speaker.}
\label{fig:comp_mlp_lstm_blstm}
\end{figure}

It is noted that LSTM and BLSTM do not perform as well as MLP on all three types of BNF representations. 
% {\color{blue} 
Experiments were carried out with different parameter settings on LSTM and BLSTM, and the system performance remained largely unchanged.
% }
% MTL-DNN modeling with both DPGMM labels and DPGMM-HMM labels. 
% Note that hyper-parameters of LSTM and BLSTM are tuned by a set of preliminary experiments, while no tuning was performed for MLP. 
Fig. \ref{fig:comp_mlp_lstm_blstm} gives the performances of HMM(P)-MUBNF learned by MLP, LSTM and BLSTM for each target language. For English (EN), different DNN structures have similar performance. For French (FR) and Mandarin (MA), the advantage of MLP over (B)LSTM is more prominent.
% As mentioned in Section \ref{subsec:exp_hybrid_modeling}, parameter tuning was performed towards LSTM and BLSTM, but not performed towards MLP.
This may be related to that the amount of training data for English is significantly greater than those for French and Mandarin.
The advantage of LSTM and BLSTM over MLP in conventional supervised acoustic modeling has been widely recognized and attributed to the capability of capturing temporal characteristics of speech. With limited training data, the benefits of recurrent structures can not be fully exploited. In our systems, contextual information is incorporated via the use of DPGMM-HMM labels and its effectiveness has been demonstrated by the experimental results.

\section{Conclusions}
\label{sec:conclusion}

BNFs learned from multilingual speech data have been proven highly effective for acoustic modeling of spoken languages. In the case of low-resource languages, the challenge of lacking transcribed data could be translated into the problem of acquiring high-quality labels to facilitate supervised DNN training. Commonly used approaches to tackling this problem include applying clustering algorithms on short-time speech frames and leveraging a language-mismatched phone recognizer to decode input speech. 
In this paper, it has been demonstrated that learning of robust BNF representations could be achieved by joint contributions from a variety of techniques, including: (1) use of speaker adapted features; (2) considering temporal dependency in speech when performing frame clustering; (3) increasing phonetic diversity by involving multiple out-of-domain languages; (4) discarding unreliable frame labels in DNN training.

The proposed methods of feature learning have been evaluated on the standard task of unsupervised subword modeling in the ZeroSpeech 2017 Challenge. The experimental results have shown that effective speaker adaptation with untranscribed training data could be achieved by using an out-of-domain ASR system. Out-of-domain ASR systems from resource-rich languages can also be utilized to provide phonetically informed labels to support multi-task learning of BNFs, in conjunction with the learning tasks based on DPGMM-HMM clustering labels. Combining different types of BNFs by vector concatenation leads to further performance improvement. The best performance achieved by our proposed system is $9.7\%$ in terms of across-speaker triphone minimal-pair ABX error rate. It is equal to the performance of the best submitted system in the ZeroSpeech 2017 and better than other recently reported systems. 

In principle, the proposed methods are expected to be effective for any combination of languages other than those in ZeroSpeech 2017. Nevertheless, our investigation has suggested that the closeness between target languages and out-of-domain languages and the amount of available training data for individual target languages might have significant impact on the goodness of learned features.

% show that: (1) Speaker adaptation based on an out-of-domain ASR is both effective and efficient; 
% (2) The alignment labels generated by DPGMM-HMM AMs provide better supervision for MTL-DNN training  than DPGMM labels; (3) The proposed frame label filtering algorithm is able to improve the reliability of DPGMM labels; (4) Concatenation of BNFs extracted by different DNNs leads to performance improvement. (5) (B)LSTM structures perform inferior to MLP {\color{blue}overall, but outperform MLP in the across-speaker and very short test condition.\st{for subword modeling in our limited training data setting}}
% Future works may focus on applying  our proposed methods to tasks such as zero-resource STD.
%\section*{Acknowledgment}
%
%%\cite{feng2016exploiting}

\ifCLASSOPTIONcaptionsoff
  \newpage
\fi


		
\bibliographystyle{IEEEtran}

\bibliography{mybib}
\end{document}